\documentclass[conference]{IEEEtran}
\IEEEoverridecommandlockouts
\usepackage{cite}
\usepackage{amsmath,amssymb,amsfonts}
\usepackage{dutchcal}
\usepackage{fontawesome}
\usepackage[linesnumbered,ruled]{algorithm2e}
\usepackage{graphicx}
\usepackage{multirow}
\usepackage{subcaption} %
\usepackage{textcomp}
\usepackage[dvipsnames]{xcolor}
\usepackage{url}
\usepackage{listings}
\usepackage{xargs}
\usepackage{booktabs}
\usepackage{makecell}
\usepackage[colorinlistoftodos, prependcaption]{todonotes}
\usepackage[hidelinks, colorlinks=false, urlcolor=blue]{hyperref}
\usepackage{cprotect}
\usepackage{circledsteps}
\usepackage{annotate-equations}
\def\BibTeX{{\rm B\kern-.05em{\sc i\kern-.025em b}\kern-.08em
    T\kern-.1667em\lower.7ex\hbox{E}\kern-.125emX}}
\usepackage{caption} 
\SetCommentSty{mycommfont}
\usepackage{xparse}
\usepackage[]{mdframed}
\usepackage{enumitem}

\tikzset{annotate equations/text/.style={font=\bfseries\footnotesize}}
\NewDocumentCommand{\code}{v}{%
\texttt{\textcolor{black}{#1}}%
}

\newcommand\encircle[1]{%
  \tikz[baseline=(X.base)] 
    \node (X) [draw, shape=circle, inner sep=0, text=black] {\strut #1};%
}

\makeatletter
\newcommand{\linebreakand}{%
  \end{@IEEEauthorhalign}
  \hfill\mbox{}\par
  \mbox{}\hfill\begin{@IEEEauthorhalign}
}
\makeatother

\begin{document}

\title{LLAMP: Assessing Network Latency Tolerance of HPC Applications with Linear Programming}

\author{
\IEEEauthorblockN{Siyuan Shen}
\IEEEauthorblockA{\textit{Department of Computer Science} \\
\textit{ETH Z\"urich}\\
Z\"urich, Switzerland \\
siyuan.shen@inf.ethz.ch}
\and
\IEEEauthorblockN{Langwen Huang}
\IEEEauthorblockA{\textit{Department of Computer Science} \\
\textit{ETH Z\"urich}\\
Z\"urich, Switzerland \\
langwen.huang@inf.ethz.ch}
\and
\IEEEauthorblockN{Marcin Chrapek}
\IEEEauthorblockA{\textit{Department of Computer Science} \\
\textit{ETH Z\"urich}\\
Z\"urich, Switzerland \\
marcin.chrapek@inf.ethz.ch}
\linebreakand
\IEEEauthorblockN{Timo Schneider}
\IEEEauthorblockA{\textit{Department of Computer Science} \\
\textit{ETH Z\"urich}\\
Z\"urich, Switzerland \\
timo.schneider@inf.ethz.ch}
\and
\IEEEauthorblockN{Jai Dayal}
\IEEEauthorblockA{\textit{SAIT Systems Architecture Lab} \\
\textit{Samsung Semiconductor}\\
San Jose, USA \\
jai.dayal@samsung.com}
\and
\IEEEauthorblockN{Manisha Gajbe}
\IEEEauthorblockA{\textit{SAIT Systems Architecture Lab} \\
\textit{Samsung Semiconductor}\\
San Jose, USA \\
m.gajbe@samsung.com}
\linebreakand
\IEEEauthorblockN{Robert Wisniewski}
\IEEEauthorblockA{\textit{SAIT Systems Architecture Lab} \\
\textit{Samsung Semiconductor}\\
San Jose, USA \\
r.wisniewski@samsung.com}
\and
\IEEEauthorblockN{Torsten Hoefler}
\IEEEauthorblockA{\textit{Department of Computer Science} \\
\textit{ETH Z\"urich}\\
Z\"urich, Switzerland \\
torsten.hoefler@inf.ethz.ch}
}

\maketitle
\IEEEpeerreviewmaketitle
\thispagestyle{plain}
\pagestyle{plain}

\begin{abstract}
The shift towards high-bandwidth networks driven by AI workloads in data centers and HPC clusters has unintentionally aggravated network latency, adversely affecting the performance of communication-intensive HPC applications. As large-scale MPI applications often exhibit significant differences in their network latency tolerance, it is crucial to accurately determine the extent of network latency an application can withstand without significant performance degradation. Current approaches to assessing this metric often rely on specialized hardware or network simulators, which can be inflexible and time-consuming. In response, we introduce LLAMP, a novel toolchain that offers an efficient, analytical approach to evaluating HPC applications' network latency tolerance using the LogGPS model and linear programming. LLAMP equips software developers and network architects with essential insights for optimizing HPC infrastructures and strategically deploying applications to minimize latency impacts.
Through our validation on a variety of MPI applications like MILC, LULESH, and LAMMPS, we demonstrate our tool's high accuracy, with relative prediction errors generally below 2\%. Additionally, we include a case study of the ICON weather and climate model to illustrate LLAMP's broad applicability in evaluating collective algorithms and network topologies.
\end{abstract}

\begin{IEEEkeywords}
Network latency tolerance, linear programming, MPI applications, high-performance computing
\end{IEEEkeywords}

\section{Introduction}

\begin{figure}[!t]
    \includegraphics[width=1\linewidth]{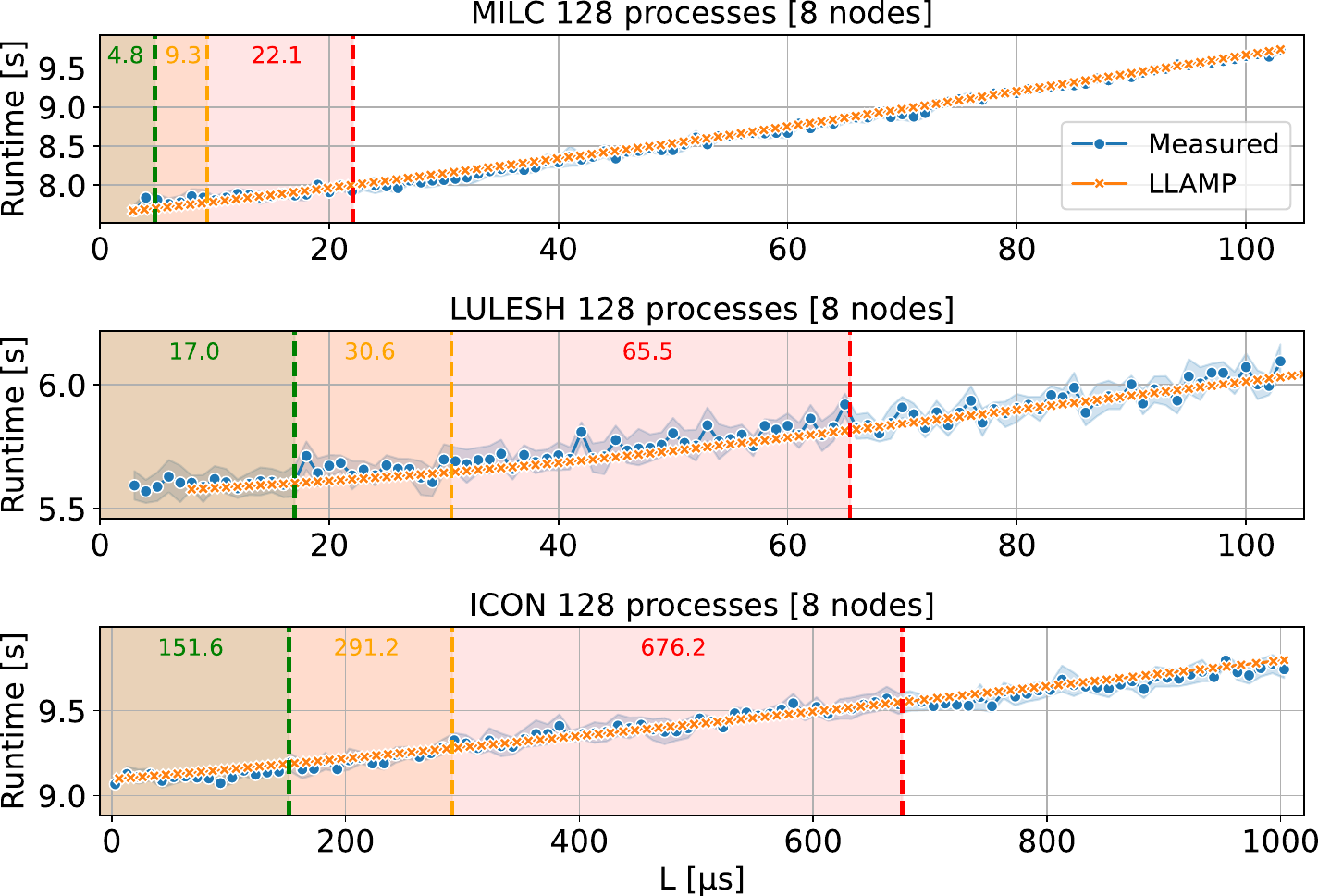}
    \cprotect\caption{An example demonstrating varying degrees of network latency tolerance among traditional HPC applications, namely MILC, LULESH, and ICON. The green, orange, and red zones correspond to the maximum network latencies before observing a performance degradation of \textcolor{ForestGreen}{1\%}, \textcolor{orange}{2\%}, and \textcolor{red}{5\%}, respectively. The comparison between measured and predicted runtime showcases the predictive accuracy of our toolchain. The tolerance intervals are calculated directly by our tool.}
    \label{fig:intro-lat-buffer}
\end{figure}

\begin{figure*}[!t]
    \centering
    \includegraphics[width=1\linewidth]{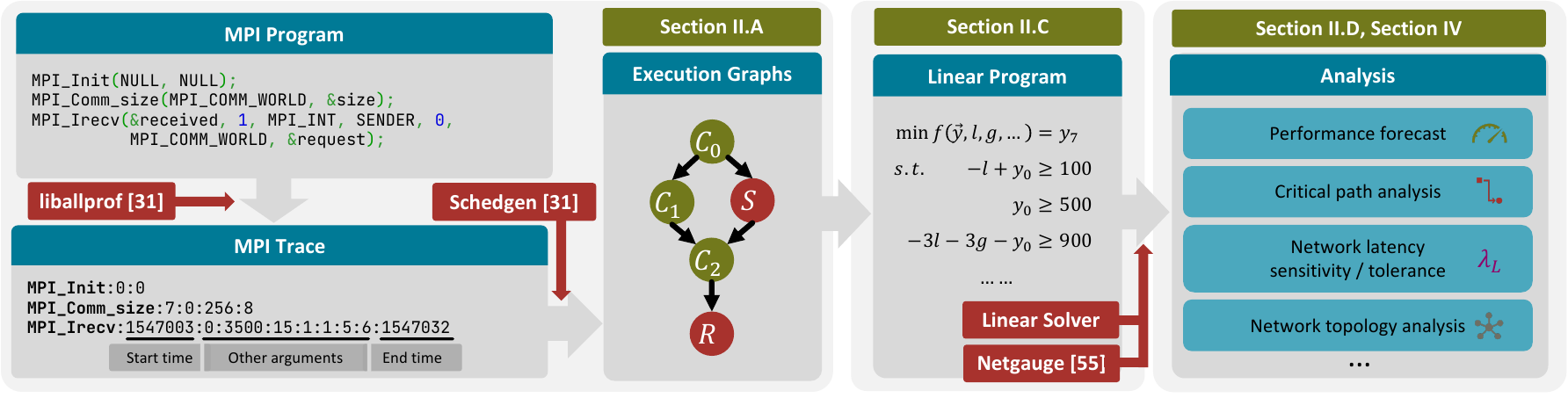}
    \caption{High-level overview of the LLAMP toolchain.}
    \label{fig:llamp_overview}
\end{figure*}

The growing demand for training large deep learning models has spurred the construction of advanced AI-focused data centers and supercomputing clusters, such as Meta's \$800 million data center ~\cite{meta_ai_center_wagner} and the upcoming Alps cluster at the Swiss National Supercomputing Center~\cite{cscs_alps}. Driven by the needs of the AI industry, there have been notable improvements in cloud hardware, which makes cloud platforms increasingly appealing for running high-performance computing (HPC) applications due to the potential cost benefits over traditional on-premise HPC clusters~\cite{cloud_computing_close_gap_guidi, cost_effective_hpc_carlyle}. However, numerous studies~\cite{noise_in_the_clouds_de_sensi, evaluation_of_hpc_applications_gupta, a_comparative_study_marathe, performance_analysis_of_hpc_exposito, running_hpc_tomic} have shown, executing large HPC applications, particularly those that are communication-intensive, can result in a suboptimal cost-performance ratio when migrated to the cloud, primarily due to increased network latency that often exceeds $10\:\mu s$~\cite{noise_in_the_clouds_de_sensi}.

Moreover, the parallel training of large deep learning models, which relies more on bandwidth than latency~\cite{deep_learning_training_naumov, varuna_athlur, architectural_requirements_ibrahim, running_hpc_tomic, using_ml_to_model_barrachina}, has prompted a rapid increase in network bandwidth. To accommodate this need, the network bandwidth has been growing exponentially in recent years~\cite{datacenter_ethernet_hoefler}. This growth, accelerated by the adoption of higher frequencies and intricate signaling techniques like PAM4, has simultaneously led to higher bit error rates (BER) in transceivers. Consequently, this requires the implementation of complex forward error correction (FEC) mechanisms, as seen with the upcoming 800G and 1.6T IEEE P802.3df standards~\cite{ieee_dambrosia}. While existing fast FEC manages to keep latency around $50\:\mathrm{ns}$, more complex FEC schemes in the future are expected to increase the decoding latency by more than $100\:\mathrm{ns}$~\cite{100_gbs_ethernet_fec_liu, fec_for_400g_bates}, subsequently adding several hundred nanoseconds to the per-link latency.

With data centers and HPC clusters evolving towards high-bandwidth networks to conform with the need of the AI sector~\cite{hammingmesh_hoefler, scalable_deep_learning_mayer}, the trade-off between bandwidth and latency becomes more pronounced, and the implications of FEC-induced latency become a concern for the design and optimization of both HPC systems and applications. Addressing this latency issue is especially important in fields where the time to solution is crucial, such as weather forecasting simulations for climate change analysis or molecular dynamics simulations for in-depth exploration of COVID-19 at the molecular level~\cite{datacenter_ethernet_hoefler}.

In large-scale MPI applications, unique communication and computation patterns inherent to each application lead to significant variations in their network latency tolerance. For instance, in Fig.~\ref{fig:intro-lat-buffer}, we present the differences in network latency tolerance across three traditional HPC applications: MIMD Lattice Computation (MILC)~\cite{milc_bernard}, LULESH~\cite{lulesh_2_karlin}, and Icosahedral Nonhydrostatic Weather and Climate Model (ICON)~\cite{icon_zangl}. The plots illustrate the impact of increasing network latency on each application's runtime. MILC exhibits the lowest tolerance to network latency, indicating that as little as $20\:\mu s$ can adversely affect its performance. Conversely, ICON demonstrates the highest latency tolerance, able to withstand more than $650\:\mu s$ of network latency before performance degradation becomes apparent. These differences emphasize the importance of network configurations to suit the specific latency tolerance profiles of each application.

Given these considerations, it is imperative to determine the extent of network latency an application can withstand without significant performance degradation. Understanding this threshold is key to designing applications that are both resilient and efficient, even under suboptimal network conditions.

However, existing methods for evaluating network latency tolerance have notable limitations. These strategies typically rely on one of three approaches, each with inherent drawbacks. The first method involves constructing elaborate performance models, demanding an in-depth understanding of the applications’ communication and computation behaviors~\cite{a_look_kerbyson, performance_modeling_milc_bauer}. This requires extensive knowledge and expertise, making it inaccessible to many practitioners. The second strategy hinges on artificially injecting latency into network communication to observe its impact on application performance. To achieve this, one can exploit specialized hardware, which is often difficult to procure and also suffers from inflexibility~\cite{evaluating_hardware_memory_disaggregation_patke, characterizing_application_rosenthal}. The third option is to rely on packet-level network simulators to predict the behaviors of applications under various conditions without the need for physical hardware. However, these simulators often require intricate configurations and can be time-consuming to execute, especially when simulating large-scale workloads~\cite{analyzing_cost_performance_bhatele, towards_million_server_besta, simulation_based_xu, enabling_parallel_mubarak}. A fundamental issue with the last two methods is the need for exhaustive parameter sweeps to ascertain an application's latency tolerance.

In response to these challenges, we introduce \emph{LLAMP} (\underline{L}ogGPS and \underline{L}inear Programming based \underline{A}nalyzer for \underline{M}PI \underline{P}rograms), a toolchain designed for efficient analysis and quantification of network latency sensitivity and tolerance in HPC applications. An overview of LLAMP is presented in Fig.~\ref{fig:llamp_overview}. By leveraging the LogGOPSim framework~\cite{loggopsim_hoefler}, LLAMP records MPI traces of MPI programs and transforms them into execution graphs. These graphs, through the use of the LogGPS model, are then converted into linear programs. They can be solved rapidly by modern linear solvers, allowing us to efficiently gather valuable metrics, such as the predicted runtime of programs, and critical path metrics. These insights are crucial for identifying an application's network latency tolerance and its performance variability under diverse network configurations. LLAMP's versatility is further demonstrated through a case study of the ICON climate model, exploring the effects of collective algorithms and network structures.

Our code is available at \url{https://github.com/spcl/llamp}.

The primary contributions of this work are as follows:
\begin{enumerate}[leftmargin=1em]
\item We derive a novel analytical model that quantifies the network latency sensitivity and tolerance of MPI applications. This model leverages linear programming, execution graphs generated from application traces, and the LogGPS model, providing a mathematical foundation for our analysis.
\item We develop LLAMP, an open-source toolchain that allows us to efficiently forecast the performance of MPI applications and compute their network latency sensitivity and tolerance. This empowers architects to tailor infrastructure designs to application needs and enables software developers to make informed decisions regarding application deployment and optimization for reduced latency sensitivity.
\item We built a software-based latency injector capable of emulating flow-level network latency with high portability, facilitating large-scale latency injection experiments without specialized hardware or administrative privileges. Using this latency injector, we validate our model on a number of MPI applications from various domains, such as LULESH, HPCG, and MILC, among others. Furthermore, we demonstrate LLAMP's broad applicability by conducting a case study on ICON, examining how collective algorithms and network topologies influence its performance.
\end{enumerate}

\section{LLAMP Toolchain}

\subsection{MPI Execution Graphs}
\label{sec:execution-graphs}
In parallel computing, analyzing MPI programs through the lens of graphs proves to be a natural and insightful methodology. As MPI facilitates parallel execution by decomposing computational tasks into distinct units, dependency graphs provide a suitable representation of the relationships and communication patterns between these tasks. As exemplified in various other works~\cite{extracting_critical_path_schulz, scalable_critical_path_bohme, investigating_dependency_graph_pereira, loggopsim_hoefler, characterizing_the_influence_hoefler}, this approach has been validated and employed successfully.

We chose to base LLAMP on the work of Hoefler et al.~\cite{loggopsim_hoefler, characterizing_the_influence_hoefler} and the reason is twofold. Firstly, the tracer and schedule generator included in the LogGOPSim toolchain allows us to easily generate execution schedules of MPI programs.
Secondly, LogGOPSim stands out as a well-tested open-source framework with an accessible codebase~\cite{using_simulation_levy, what_if_mpi_collective_riesen}, facilitating extensibility and customization to suit our specific needs.

To elaborate, the schedule generator, \textbf{Schedgen}, parses the traces collected from the light-weight tracing library, \textbf{liballprof}, and converts them into execution graphs that capture the precedence of events and task dependencies in GOAL format~\cite{group_operation_assembly_language_hoefler}. An MPI execution graph is a directed acyclic graph (DAG) that contains three vertex types, namely \textit{send}, \textit{recv}, and \textit{calc}, which denote point-to-point (p2p) sends and receives and computations, respectively. Edges connecting \textit{send} and \textit{recv} pairs are termed \textit{communication edges}.
By exploiting the difference in timestamps of consecutive MPI operations, Schedgen infers the amount of computation that occurred. In addition, Schedgen is able to substitute collective operations with p2p algorithms based on user specifications (see \cite{loggopsim_hoefler} for more details).

The LogGOPS model, a derivative of the LogP family~\cite{logp_culler, a_survey_rico_gallego}, serves to quantify the communication costs in parallel applications. In this model, $L$ denotes the maximum latency between two processors, which is the network latency we will focus on in this work. Parameter $o$ represents the overhead on the CPU per message, and $g$ is the gap between two consecutive messages. $G$, as introduced by Alexandrov et al.\cite{loggp_alexandrov}, models the gap per byte of a message, and it is equivalent to the inverse of bandwidth. Additionally, $O$ is the CPU overhead per byte, and $P$ represents the number of processes. The parameter $S$ sets the size threshold for employing the \emph{rendezvous protocol}, where messages smaller than $S$ are sent immediately (i.e., the eager protocol), whereas larger messages require synchronization between sender and receiver before transmission. As highlighted by Hoefler et al.~\cite{characterizing_the_influence_hoefler}, $O$ is commonly negligible due to high overlap. Therefore, we adopted the LogGPS model for our analysis.

\begin{figure}[!t]
\centering
\includegraphics[width=1\linewidth]{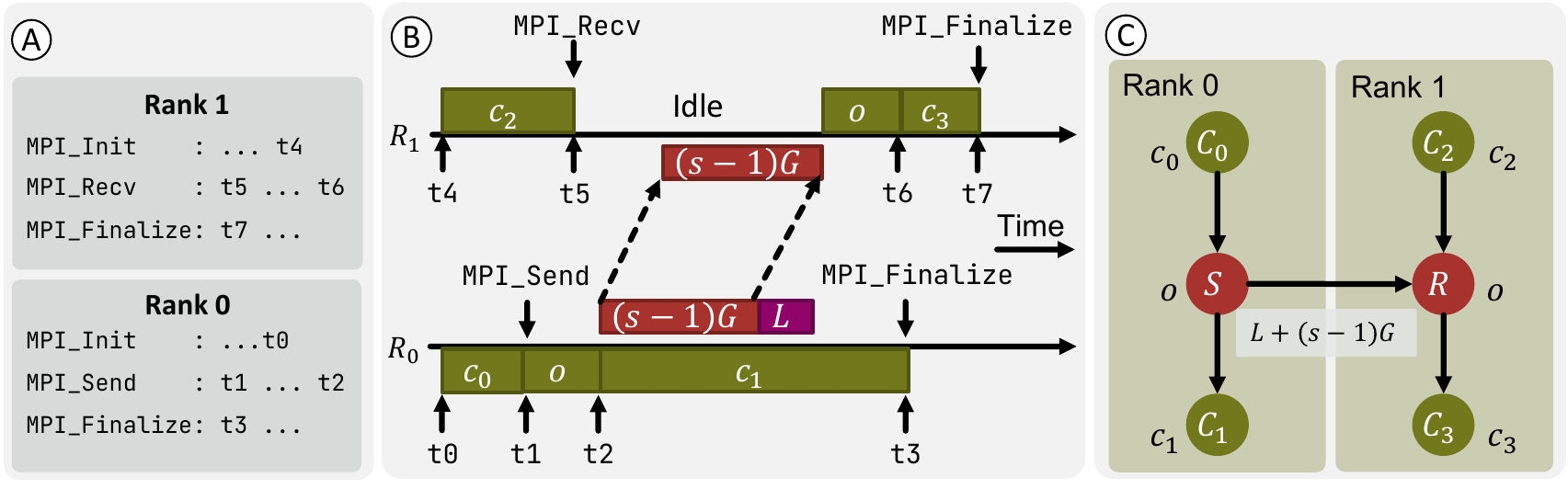}
\cprotect\caption{An example illustrating the transformation of blocking p2p operations into an execution graph, assuming that the eager protocol is used. \protect\encircle{A} lists collected traces with only the start and end timestamps. \protect\encircle{B} shows the corresponding space-time diagram. In \protect\encircle{C}, \emph{calc} vertices are marked in green while \emph{send} and \emph{recv} vertices are in red.}
\label{fig:dependency-schedule-example}
\end{figure}

In panel \encircle{A} of Fig.~\ref{fig:dependency-schedule-example}, we present the trace of a simple MPI program.
Moving to panel \encircle{B}, the cost of \emph{calc} vertices are inferred from the start and end timestamps for each operation in the trace. For instance, the interval $c_0 = t0 - t1$ on rank 1 indicates a computation period of $c_0$ before initiating \code{MPI_Send}.  Through this process, Schedgen determines the computational workload and the operational dependencies within the application. Subsequently, by assigning specific LogGPS parameters to the sends and receives, we can effectively model an application's behavior across different network configurations. For an illustration of this approach applied to nonblocking communications, refer to Fig.~\ref{fig:dependency-schedule-nonblocking-example} in the appendix.

\begin{figure*}
\begin{subfigure}{.23\textwidth}
\centering
\includegraphics[height=3.2cm]{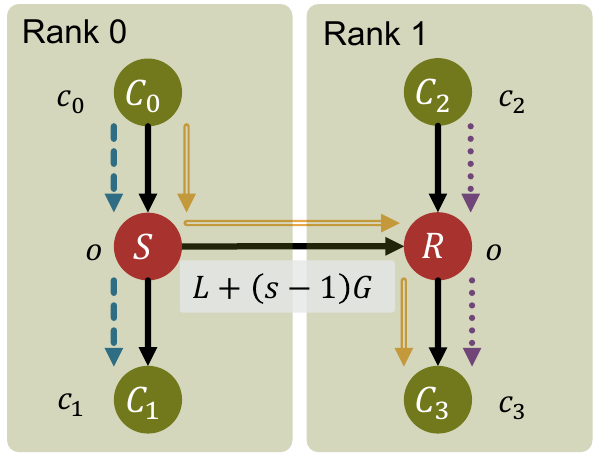}
\caption{Three possible paths through the DAG are highlighted with arrows of different patterns.}
\label{fig:sensitivity-example-base}
\end{subfigure}
\hspace{0.01\textwidth}%
\begin{subfigure}{.33\textwidth}
\centering
\includegraphics[height=3.2cm]{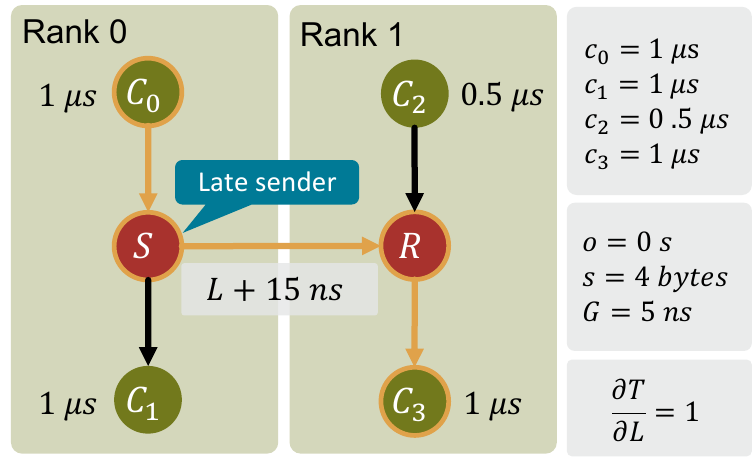}
\caption{Values are assigned to parameters in a way that makes $\partial T/\partial L = 1$. The critical path is highlighted in orange.}
\label{fig:sensitivity-example-1}
\end{subfigure}
\hspace{0.01\textwidth}%
\begin{subfigure}{.43\textwidth}
\centering
\includegraphics[height=3.2cm]{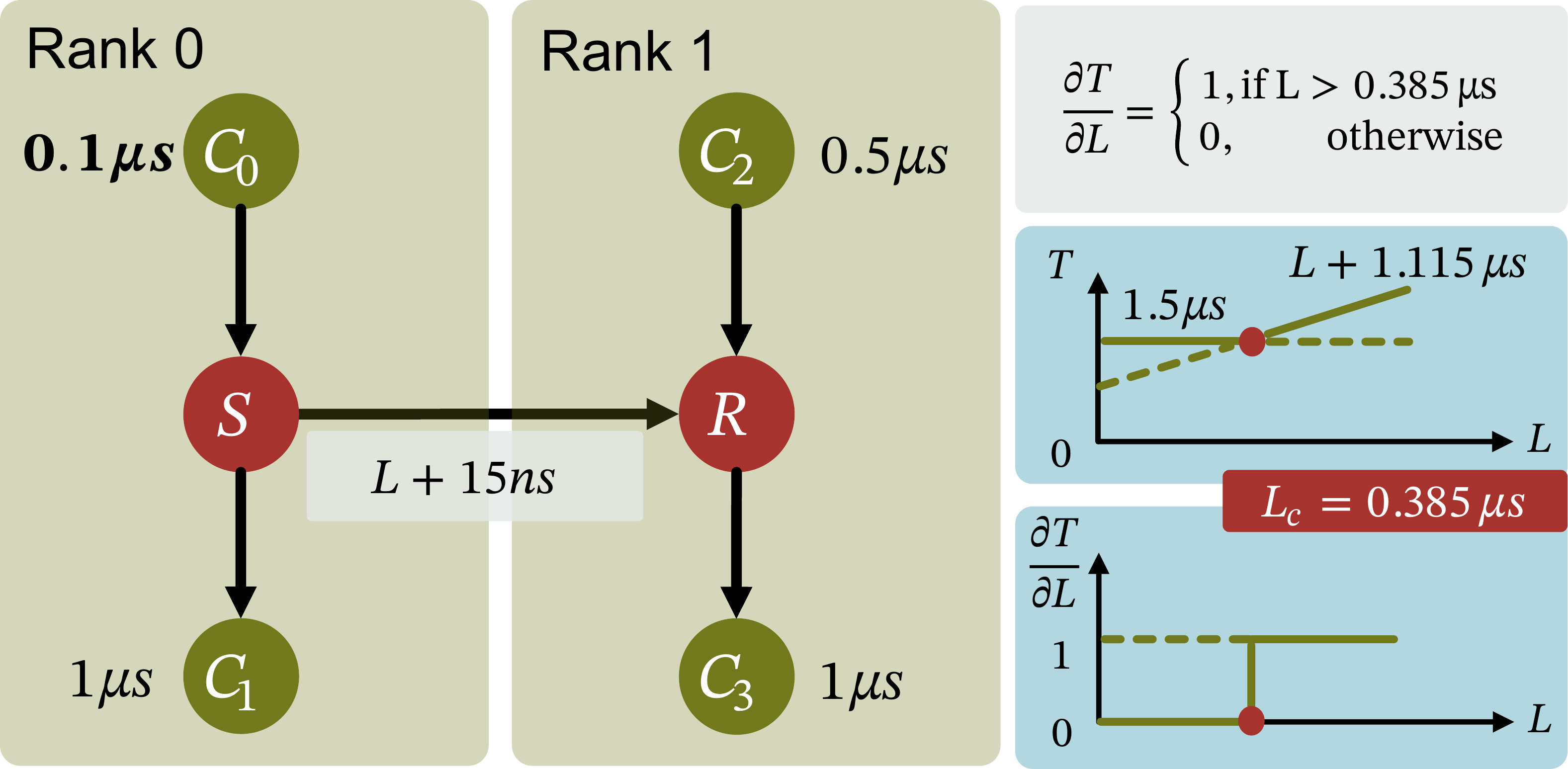}
\caption{If $t_0$ is reduced to $0.1\mu s$, $\partial T/\partial L$ will be dependent on the value of $L$. The relationship between $T$, $\partial T / \partial L$, and $L$ are plotted. The critical latency $L_c$ is calculated to be $0.385 \mu s$.}
\label{fig:sensitivity-example-2}
\end{subfigure}
\caption{An example demonstrating that the network latency sensitivity of a program is determined by the number of messages along the critical path of the graph and is also dependent on the value of parameter $L$ itself.}
\label{fig:sensitivity-example}
\end{figure*}

\subsection{Sensitivity Analysis}
\label{sec:sensitivity-analysis}
In this work, we explore the concept of network latency sensitivity. To frame our discussion, we will begin with an overview of sensitivity analysis (SA) formally.
SA examines how a set of $N$ input variables $x = \{ x_1, \ldots x_N \}$ influences the output $y = \{ y_1, \ldots, y_D \}$ of $y = g(x)$ where $g: \mathbb{R}^N \rightarrow \mathbb{R}^D$~\cite{sensitivity_analysis_borgonovo, sensitivity_analysis_saltelli}.
When $g$ is differentiable, derivative-based local SA can be performed by computing the partial derivative of $y$ w.r.t. the $i$-th value in the input, $x_i$, denoted as $\lambda_i = \frac{\partial y}{\partial x_i}\Bigr|_{x_0}$, where $\lambda_i$ is the sensitivity measure of $x_i$, and $x_0 \in \mathbb{R}^n$ is the fixed point for evaluating the derivative~\cite{sensitivity_analysis_of_model_output_borgonovo, sensitivity_analysis_of_environmental_models_pianosi}.

We start by exploring the impact of network latency, $L$, on the runtime of MPI programs. According to derivative-based local SA, we can express the network latency sensitivity as
\begin{align}
\footnotesize
\lambda_L(\mathcal{G}, \theta) = \partial T(\mathcal{G}, \theta) / \partial L
\end{align}
where $\mathcal{G}$ represents an execution graph, $\theta$ denotes a specific configuration (i.e., a vector defining each parameter's value in LogGPS), and $T(\mathcal{G}, \theta)$ computes the program runtime given $\mathcal{G}$ and under the configuration $\theta$. $\mathcal{G}$ and $\theta$ are omitted if they are unambiguous. Intuitively, $\lambda_L$ characterizes \textit{the variation in a program's runtime if $L$ is increased by 1 unit}.

In a parallel program, its runtime is determined by the critical path~\cite{extracting_critical_path_schulz, scalable_critical_path_bohme, critical_path_candidates_chen}. Therefore, to compute $\lambda_L$, it is crucial to derive an expression for the critical path as a function of $L$.
To illustrate this, we refer to Fig.~\ref{fig:sensitivity-example}. Fig.~\ref{fig:sensitivity-example-base} contains the execution graph for two ranks, where rank 0 executes \code{MPI_Send} while rank 1 executes \code{MPI_Recv}. Both ranks perform computation before and after the MPI operations, indicated by the green vertices connected to nodes $S$ and $R$ with costs $c_0$ through $c_3$, respectively. There are in total three possible paths through the DAG, and without assigning values to the costs and the LogGPS parameters, the critical path can be any of them. Hence, $T$ can be expressed formally as

\vspace{.6em}
\begin{align}\label{eq:sensitivity-example}
\footnotesize
\begin{split}
T(\theta) = \max ( &\eqnmarkbox[NavyBlue]{p1}{t_0 + o + t_1}, \eqnmarkbox[Purple]{p3}{t_2 + t_3 + o},\\
                   &\eqnmarkbox[Orange]{p2}{t_0 + o + L + (s - 1) G + t_3 + o})
\end{split}
\end{align}
\annotate[yshift=.8em,xshift=1em]{above, left}{p1}{$C_0 \rightarrow S \rightarrow C_1$}
\annotate[yshift=-0.5em]{below, right}{p2}{$C_0 \rightarrow S \rightarrow R \rightarrow C_3$}
\annotate[yshift=.8em]{above, right}{p3}{$C_2 \rightarrow R \rightarrow C_3$}
\vspace{1em}

where each term in the \textit{max} operator represents a distinct path. In Fig.~\ref{fig:sensitivity-example-1}, we show that after substituting $c_0 = c_1 = c_3 = 1\:\mu s$, $c_2 = 0.5\:\mu s$, $s = 4$, $o = 0\:s$, and $G = 5\:\mathrm{ns}$ into Equation~\ref{eq:sensitivity-example}, it can be simplified to $T = L + 2.015\:\mathrm{\mu s}$, which leads to the result $\lambda_L = \partial (L + 2.015)/\partial L = 1$. This is designed to highlight that depending on vertices' costs and their precedence in the graph, some communication edges will stay on the critical path regardless of the value of $L$. In this scenario, due to a late sender, the edge $S \rightarrow R$ remains on the critical path, making $\lambda_L$ independent of the $L$'s value. 

Nonetheless, if $c_0$ is changed from $1 \mathrm{\mu s}$ to $0.1 \mathrm{\mu s}$ while all other variables stay the same, Equation~\ref{eq:sensitivity-example} will evaluate to: $T = \max(L + 1.115\:\mathrm{\mu s}, 1.5\:\mathrm{\mu s})$, in which case the result yielded by the \textit{max} operator, and consequently, the critical path, will be dependent on $L$. After plotting $T = 1.5\:\mu s$ and $T = L + 1.115\:\mu s$ as shown in Fig.~\ref{fig:sensitivity-example-2}, we notice that when $L \leq 0.385\:\mu s$, $T$ stays constant as the communication cost is overlapped. Once $L$ is larger than this threshold, $T$ starts to increase linearly. We refer to this point at which the critical path changes the \textit{critical latency}, denoted as $L_c$. The same overlapping effect can also be observed within a rank when nonblocking operations are executed.

To generalize, the runtime $T$ of an MPI program under the LogGPS model can be expressed as:
\begin{align}\label{eq:latency-sensitivity}
\footnotesize
T(\mathcal{G}, \theta) = \max (a_0 L + C_0, a_1 L + C_1, \ldots, a_n L + C_n)
\end{align}
where the $i$-th term in \textit{max} is the cost of a distinct path after simplification, $a_i$ is the number of communication edges along the path and $C_i$ is a constant representing all other costs. If two paths share the same $a_i$, the one with larger $C_i$ will be kept. From Equation~\ref{eq:latency-sensitivity}, we know that $n$ is bounded by the length of the longest chain of messages in the graph. 

We gained two insights from the formulation of $T$. Firstly, 
\emph{the number of messages along the critical path dictates $\lambda_L$}. Secondly, \emph{the value of $L$ has a second-order effect on the network latency sensitivity}. As $L$ increases, more communication edges that cannot be overlapped will lead to an increase in $\lambda_L$, and less latency tolerant a program will be. Therefore, it is crucial for network engineers to understand both the critical latencies and $\lambda_L$ of MPI programs within an interval of interest. With this knowledge, they can make informed decisions regarding network configurations, improving program performance and resilience to varying levels of $L$.

\subsubsection{Generalization}

Extending our sensitivity analysis beyond network latency, $L$, we can apply similar principles to assess the impact of other parameters, such as $G$ in LogGPS. For this purpose, consider the following expression:
\begin{align}\label{eq:bandwidth-sensitivity}
\footnotesize
T(\theta) = \max ( s_0 G + C_0, s_1 G + C_1, \ldots, s_m G + C_m)
\end{align}
Similar to Equation \ref{eq:sensitivity-example}, each term in \textit{max} represents the cost of a path after simplification. In this case, $s_i$ is approximately the number of bytes contained in messages along each path. Therefore, after evaluation, the \emph{bandwidth sensitivity} measure $\lambda_G(\mathcal{G}, \theta) = \frac{\partial T(\mathcal{G}, \theta)}{\partial G}$ can be seen as reflecting the total message size along the critical path. In essence, since $T$ is determined by the critical path, as long as we can express the cost of each path as a function of a parameter, we can derive the sensitivity measure for that parameter.
Despite this generalization, our investigation will focus on exploring the impact of $L$.

\subsection{Linear Programming in LLAMP}

Sensitivity measures of parameters can be computed through two conventional graph analysis approaches. 
The first approach involves traversing the graph twice: first to assign timestamps to all vertices for a specific $\theta$, and second to obtain the critical path and relevant metrics. Despite its linear time complexity (i.e., $O(|E| + |V|)$), it requires parameter sweeps to identify critical latencies. The second approach aims to find the general expression for $T$ directly, such as Equations~\ref{eq:latency-sensitivity} and~\ref{eq:bandwidth-sensitivity}, enabling us to determine all critical latencies at once. Yet, since this requires iterating over all possible paths, it is generally intractable. An alternative would be using dynamic programming (DP) to store path-related information as a map for each vertex. For $\lambda_L$, the time complexity would be $O(n|E| + |V|)$, where $n$ is the length of the longest message chain in the graph. While DP seems effective, when we tested it on a graph of LULESH with around 500K vertices, it took more than 4 hours to finish, which indicates that DP is also not scalable, even for graphs of moderate sizes.

\subsubsection{Linear Program Formulation}

While conventional graph analysis approaches for computing sensitivity measures exhibit inherent limitations, we present an alternative methodology that employs linear programming (LP). LP is a technique that maximizes or minimizes a linear function known as the \emph{objective function} subject to a set of linear constraints~\cite{linear_programming_britannica}:
\begin{align*}
\text{Maximize:} \quad & z = \mathbf{c}^T \mathbf{x} & & z \in \mathbb{R}, \mathbf{c} \in \mathbb{R}^n, \mathbf{x} \in \mathbb{R}^n \\
\text{subject to:} \quad & \mathbf{A}\mathbf{x} \leq \mathbf{b}, \mathbf{x} \geq 0 & & \mathbf{A} \in \mathbb{R}^{m \times n}, \mathbf{b} \in \mathbb{R}^m
\end{align*}
The canonical formulation of an LP, as illustrated above, involves $\mathbf{A}$, $\mathbf{b}$, and $\mathbf{c}$, which all contain constants defined by the restrictions of the given problem. A solution $\mathbf{x}$ is \emph{feasible} if it satisfies all the constraints. The goal of solving an LP is to determine the values of \emph{decision variables} in vector $\mathbf{x}$ that lead to the optimal solution in the \emph{feasible region}. A \emph{basis} is a subset of $\mathbf{Ax} \leq \mathbf{b}$, $\mathbf{x} \geq 0$ containing $n$ linearly independent constraints. Geometrically, it uniquely determines a vertex on the polyhedron shaped by the feasible region. After solving an LP, a constraint is considered \emph{tight} if the solution satisfies the constraint with equality, essentially defining a boundary of the polyhedron where the optimal solution is located.

To illustrate the relationship between execution graphs and LP, we will examine the running example presented in Fig.~\ref{fig:sensitivity-example-2}. By substituting the values of parameters into Equation~\ref{eq:latency-sensitivity} and re-organizing it, we gain the following formula:

\begin{align}\label{eq:lp-max-example}
\footnotesize
T = \max ( 
\eqnmarkbox[NavyBlue]{p1}{1.1},
\max (\eqnmarkbox[Orange]{p2}{L + 0.115}, \eqnmarkbox[Purple]{p3}{0.5}) + 
\eqnmarkbox[Green]{vertex5}{1}
)
\end{align}
\annotate[yshift=.7em]{above, left}{p1}{$C_0 \rightarrow S \rightarrow C_1$}
\annotate[yshift=.7em]{above, left}{p2}{$C_0 \rightarrow S \rightarrow R$}
\annotate[yshift=.7em, xshift=.5em]{above, left}{p3}{$C_2 \rightarrow R$}
\annotate[yshift=.7em]{above, left}{vertex5}{$C_3$}
\vspace{-1em}

In Equation~\ref{eq:lp-max-example}, the costs of paths or vertices associated with the terms are labeled accordingly. To convert this expression to an LP problem, each \textit{max} operator can be interpreted as two constraints where each constraint corresponds to one term, and the latency $L$ needs to be replaced with a decision variable. The linear program, after conversion, is shown as follows:
\begin{equation}
\label{eq:lp-model-example}
\begin{gathered}
\min_{\mathcal{l}, y_1, y_2}\; t\\
\footnotesize
\begin{alignedat}{2}
\textup{s.t.}\;(1) \; y_1\geq \mathcal{l} + 0.115, \; (2) \; y_1 \geq 0.5, \; (3) \; t\geq 1.1,  \; (4) \; t \geq y_1 + 1
\end{alignedat}
\end{gathered}
\end{equation}
where $y_1$ corresponds to the outcome of the inner \textit{max} whereas $t$ represents the result of the outer \textit{max} and serves as the variable to be minimized in the objective function. Note that the LP is not formulated in its canonical form for easier understanding. To obtain $T$ for a given $L$, one only needs to add the constraint $\mathcal{l} \geq L$, and solve for the objective value. To distinguish between the decision variables denoting different LogGPS parameters and the constants that are assigned as their lower bounds, we use the lowercase letter in mathematical font to indicate the decision variables (e.g., $\mathcal{l}$, $\mathcal{g}$, $\mathcal{o}$).

\begin{figure}[!htp]
    \centering
    \includegraphics[width=.8\linewidth]{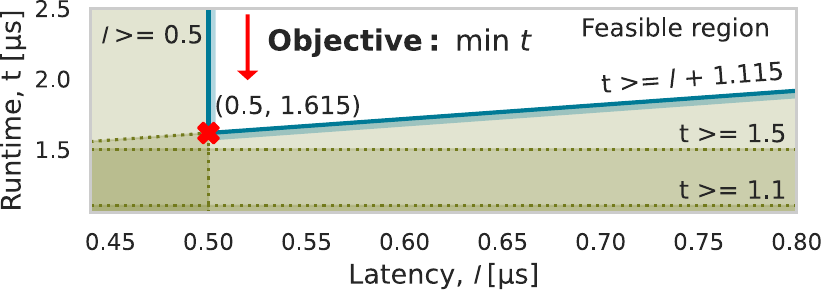}
    \caption{Visualization of Equation~\ref{eq:lp-model-example}. Dotted green lines define the linear constraints. Shaded areas mark the infeasible region. Blue lines highlight the borders of the feasible region.}
    \label{fig:lp-min-t}
\end{figure}

Fig.~\ref{fig:lp-min-t} visualizes the LP in Equation~\ref{eq:lp-model-example}. By adding $l \geq 0.5$, we aim to calculate the runtime when $L = 0.5\:\mu s$. While the LP originally includes three decision variables, we have integrated the value of $y_1$ into $(4)$, reducing the problem to two dimensions for easier visualization. The visualization clearly pinpoints the result at the coordinate $(0.5, 1.615)$, demonstrating that when $L = 0.5\:\mu s$, the runtime $T = 1.615\:\mu s$.

To generalize, we consider the start and finish times of a vertex $v$ as $t_s(v)$ and $t_e(v)$. While traversing the graph in topological order, $t_e(v) = t_s(v) + cost(v)$, where $cost(v)$ returns the execution time of $v$. If $v$ has only one predecessor, $u$, then $t_s(v) = t_e(u)$. Conversely, if $v$ has more than one predecessor, we introduce a new decision variable $y_v$ and add the following set of constraints $\{y_v \geq t_s(u) + cost(u, v)| \forall u \in V, (u, v) \in E\}$. This procedure is outlined by Algorithm~\ref{alg:lp-generation} in the appendix. Moreover, we describe how the rendezvous protocol is supported in Appendix~\ref{appendix:sec:support-for-rendezvous}. Since we are only traversing the graph once, the time complexity of this method is $O(|V| + |E|)$. The space complexity is also $O(|V| + |E|)$ as the number of variables and constraints in the LP are bounded by $|V|$ and $|E|$, respectively. The complexity of solving the LP is discussed in Section~\ref{sec:advantage-of-lp}.

\subsection{Performance Metrics}

\subsubsection{Sensitivity Measures}

After converting the execution graph into an LP, we gain the ability not only to estimate an application's runtime but also to leverage the concept of \emph{reduced cost} (RC) for assessing network latency sensitivity. The RC of a decision variable quantifies the amount at which the objective value will change if the value of the variable varies by 1 unit. As illustrated in Fig.~\ref{fig:lp-min-t}, the reduced cost for $l$ equals 1, indicating a direct correlation where an increase in $l$ by $1\:\mu s$ results in an increase in $t$ by $1\:\mu s$. Thus, the RC of $l$ directly reflects $\lambda_L$. Since by solving an LP, we automatically calculate the RC for all decision variables, we can obtain sensitivity measures of parameters such as $L$ and $G$ much more efficiently, unlike graph analysis, which requires two separate traversals of the graph to deduce the same metric.

Constraints in LP problems can be viewed as an alternative representation of edges in graphs. Intuitively, \emph{if a set of constraints are tight after optimization, their corresponding edges are on the critical path}. For instance, when we introduce $\mathcal{l} \geq 0.5$ to Equation~\ref{eq:lp-model-example} and solve for $t$, we observe that constraints $(1)$ and $(4)$ are tight, indicating that the path they represent, $\textcolor{Orange}{C_0 \rightarrow S \rightarrow R \rightarrow C_3}$ in Fig.\ref{fig:sensitivity-example-base}, is the critical path. As a basis closely correlates with constraints, if the optimal basis of an LP remains constant, the critical path of its corresponding execution graph will be unchanged. Given any $L$ as the lower bound of $\mathcal{l}$, we can compute its \emph{range of feasibility}, $L_{\mathit{fl}} \leq L \leq L_{\mathit{fu}}$, where $L_{\mathit{fl}}$ and $L_{\mathit{fu}}$ are the feasibility lower bound and upper bound respectively. Within this range, both the critical path and $\lambda_L$ would remain the same. Note that a linear solver produces the range of feasibility of all variables after optimization. Algorithm~\ref{alg:critical-latencies} describes the steps to obtain critical latencies within a given range of $L$. The difference between our approach and graph analysis is that every time the LP is solved, it provides us with information about the region around the optimal basis. Thus, we can explore an interval more efficiently. Appendix~\ref{appendix:sec:computing-critical-latencies} provides a thorough example and further explanation on the algorithm.

Derived from $\lambda_L$, which indicates the number of messages along the critical path as discussed in Section~\ref{sec:sensitivity-analysis}, the product $L \cdot \lambda_L$ reveals the total time consumed by network latency on the critical path. Consequently, we define $\rho_L = T / (L \cdot \lambda_L)$, termed as the \emph{$L$ ratio}. $\rho_L$ highlights what fraction of the critical path's execution time is due to network latency. By calculating this ratio, we gain insight into the overall impact of network latency on the performance of the application.

\subsubsection{Network Latency Tolerance}
\label{sec:network-latency-tolerance}
Beyond assessing network latency sensitivity, a significant benefit of LP lies in its ability to directly calculate the network latency tolerance of applications. Importantly, in Fig.~\ref{fig:intro-lat-buffer}, different tolerance values were not derived from iterating over predicted or actual application runtimes; instead, they were directly computed by solving LPs. 

To achieve this, it requires two simple adjustments to the original LP. Firstly, we change the objective function from minimizing $t$ to maximizing $l$. Following this, an additional constraint is introduced to set the maximum allowable execution time $t$. Solving the LP under these conditions yields the precise maximum tolerable $L$, beyond which the application's runtime exceeds a predefined threshold. We thus define the \emph{$x\%\:L$ tolerance} of an application as the highest $L$ tolerable before experiencing more than $x\%$ degradation in performance.

\begin{figure}[!htp]
    \centering
    \includegraphics[width=.8\linewidth]{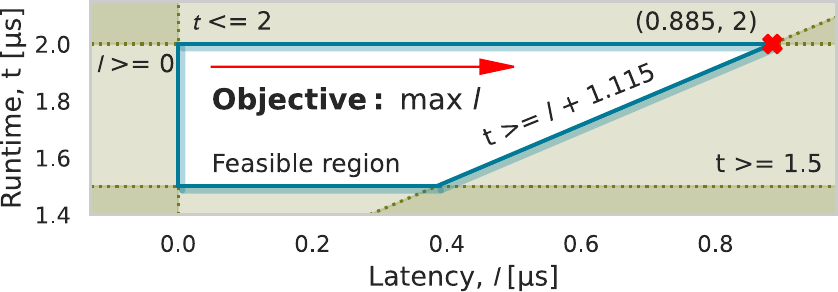}
    \caption{The alternative optimization problem that tries to maximize $l$ given an upper bound of the decision variable $t$. The shaded regions are infeasible. $t \geq 1.1$ is not shown.}
    \label{fig:lp-max-l}
\end{figure}

Fig.~\ref{fig:lp-max-l} showcases the modified LP applied to our running example. Our goal here is to determine the maximum network latency, $L$, that keeps the application's runtime below $2\:\mu s$. Hence, the model incorporates a constraint $t \leq 2\:\mu s$ and optimizes for $l$. As illustrated, the LP efficiently identifies the optimal solution as $0.885\:\mu s$. This underscores LP's effectiveness as a tool for network architects and software engineers, providing a direct measure of network latency tolerance without the need for exhaustive parameter sweeps.

\subsubsection{Advantages of Linear Programming}
\label{sec:advantage-of-lp}

While transforming an execution graph into a linear program has linear time and space complexities, some doubts might be raised regarding the efficiency of this approach considering that solving a linear program is known to have exponential time complexity in the worst case with the simplex algorithm~\cite{lp_algorithms, how_good_is_the_simplex_algorithm_klee}.
In theory, one might expect traditional simulation or graph analysis methods to be more scalable. However, the reality is different.

\begin{figure}[!htp]
    \centering
    \includegraphics[width=1\linewidth]{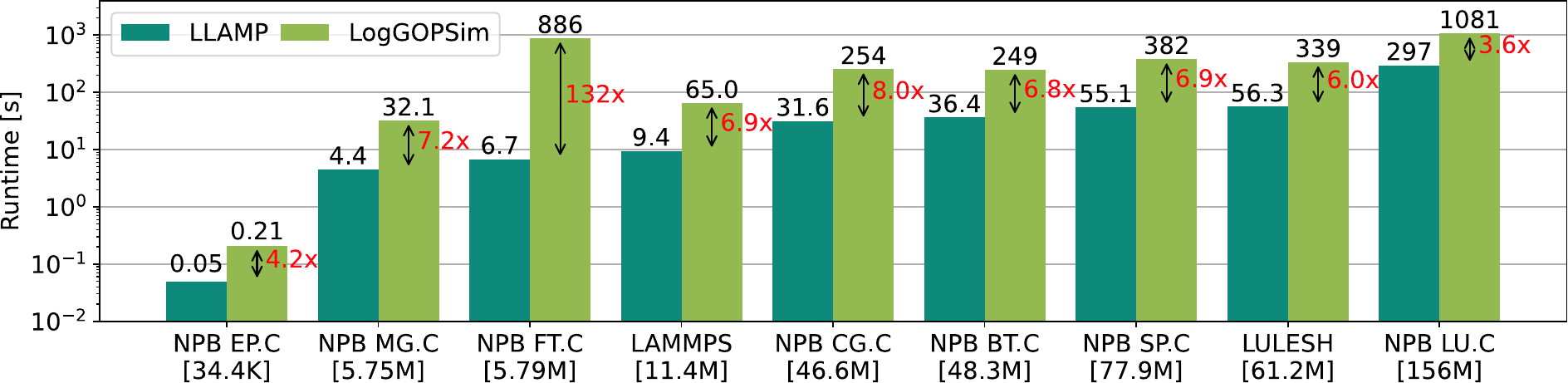}
    \caption{Runtime of the linear solver used by LLAMP vs. LogGOPSim across various applications. The number in the brackets lists the number of events in the execution graph.}
    \label{fig:lp-runtime}
\end{figure}

\begin{figure*}[!t]
\centering
\begin{subfigure}{.23\textwidth}
    \centering
    \includegraphics[height=3.2cm]{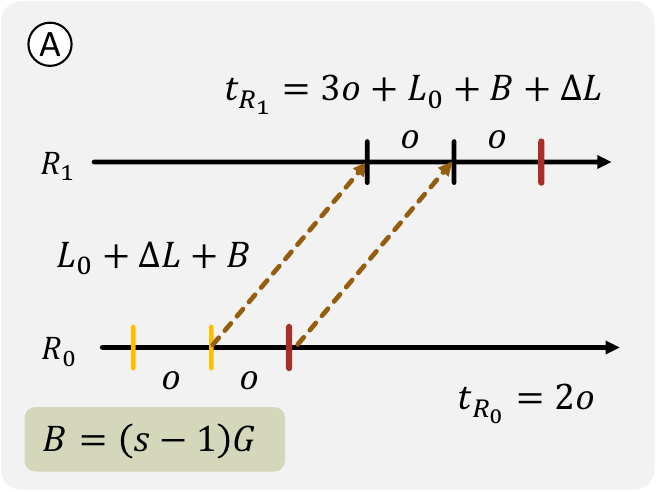}
    \caption{Intended outcome of latency injection in which $\Delta L$ latency is added to the base latency $L_0$.}
    \label{fig:latency-injector-intended}
\end{subfigure}
\hspace{0.01\textwidth}%
\begin{subfigure}{.23\textwidth}
    \centering
    \includegraphics[height=3.2cm]{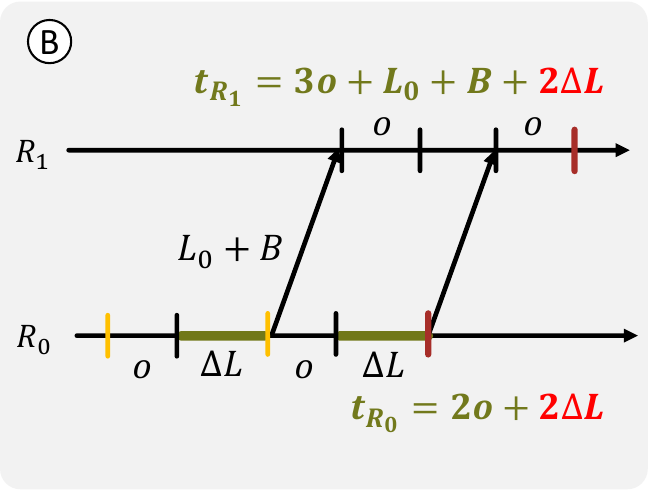}
    \caption{Adding delay to sends without any additional threads. (Underwood et al.~\cite{measuring_network_latency_underwood})}
\label{fig:latency-injector-sender}
\end{subfigure}
\hspace{0.01\textwidth}%
\begin{subfigure}{.23\textwidth}
    \centering
    \includegraphics[height=3.2cm]{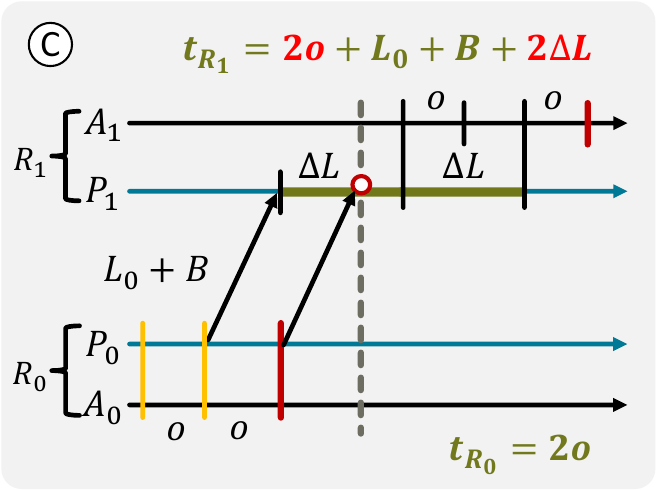}
    \caption{Adding delay on the receiver's side with a progress thread.}
    \label{fig:latency-injector-receiver}
\end{subfigure}
\hspace{0.01\textwidth}%
\begin{subfigure}{.23\textwidth}
    \centering
    \includegraphics[height=3.2cm]{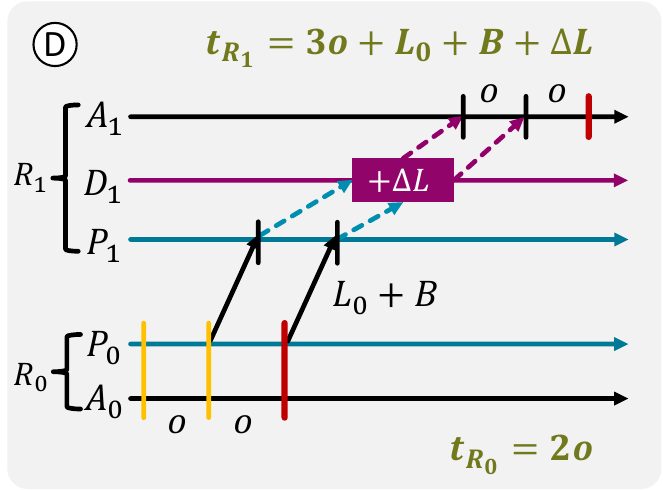}
    \caption{Adding delay to receives with progress and delay threads. (Our work)}
    \label{fig:latency-injector-delay-thread}
\end{subfigure}
\cprotect\caption{\protect\encircle{A} shows the expected result when $\Delta L$ is added to the network latency in a scenario where the sender ($R_0$) performs two consecutive eager \code{MPI_Send}s while the receiver ($R_1$) posts two \code{MPI_Recv}s prior to the sends. $t_{R_0}$ and $t_{R_1}$ represent runtime of the ranks, respectively. The orange vertical lines indicate the time at which each send starts. The arrows denote the transmission of entire messages. $L_0$ represents the network's base latency. For brevity, the bandwidth cost for message transmission, $(s-1)G$, is simplified to $B$. Subsequent panels depict various network latency injector implementations and their effects. Expressions in each panel that match those in \encircle{A} are marked in green, while those that differ are in red.}
\label{fig:latency-injector-types}
\end{figure*}

The runtime comparison between LLAMP utilizing Gurobi~\cite{gurobi}, a state-of-the-art linear solver, and LogGOPSim is presented in Fig.\ref{fig:lp-runtime}. More details of the experiment are available in Section~\ref{appendix:sec:solver-experiment} of the appendix. Our choice of LogGOPSim for comparison is based on two key factors. Firstly, LogGOPSim stands out as one of the most efficient and scalable simulators currently available~\cite{loggopsim_hoefler, towards_million_server_besta}. Secondly, the graphs used to generate the LP models in our analysis are created by Schedgen, a component also utilized by LogGOPSim, ensuring a fair comparison.
Despite LogGOPSim's notable speed and scalability, it is consistently outperformed by the linear solver often by a factor of more than $6\times$, regardless of the problem size. This discrepancy arises mainly because the presolve phase of the linear solver efficiently eliminates all redundant constraints with advanced heuristics~\cite{advanced_gurobi_algorithms_wunderling, presolving_in_lp_andersen}. This drastically reduces the solve time. Moreover, the barrier algorithm, also referred to as the interior point method~\cite{applied_mathematical_programming_bradley, lp_algorithms}, employed by the solver allows the problems to be easily parallelized. Consequently, in practice, employing LP for performance forecasting is more efficient.

\section{Validation}

As LLAMP forecasts application performance, we can artificially introduce latency to network communication so as to compare its actual impact with LLAMP's predictions. In this section, we present two major contributions: a software-based latency injector for precise, scalable flow-level latency injection without specialized hardware or admin privileges, and a demonstration of LLAMP's predictive accuracy.

\subsection{Network Latency Injector}

Emulating network latency accurately is a complex task. 
For instance, take the situation shown in Fig.~\ref{fig:latency-injector-types}, where our goal is to inject an extra delay, $\Delta L$, into the network. We expect $R_0$ to complete sending messages by timestamp $2o$, and $R_1$ to receive them at $3o + L_0 + B + \Delta L$, assuming both start at timestamp 0. The simplest method to add latency is to delay the send operation by $\Delta L$. As seen in \encircle{B}, this unintentionally delays both $R_0$ and $R_1$. Note that even with an additional progress thread to handle the sending, one \code{MPI_Send} would still have to wait until $o + \Delta L$ before returning, delaying the next send. Underwood et al.’s work~\cite{measuring_network_latency_underwood} exploited this basic approach by hooking their latency injector into the \code{post_send} function in \emph{libibverbs} library. However, as we have shown here, it would introduce unwanted delays to consecutive send operations.

The approach in \encircle{C} adds a progress thread to the receiver to process the delay, freeing the sender from the wait. Nonetheless, this method faces issues when $\Delta L$ exceeds $o$, which is often the case in practice. The progress thread is still handling the first message's delay when the second arrives. This leads to the second message being delayed to $o + L_0 + B + 2 \Delta L$, not the expected $2o + L_0 + B + \Delta L$. To concurrently process delays for multiple messages, the progress thread would need to track release times while polling the receive queue, which can greatly reduce the accuracy and resolution of $\Delta L$.

Our solution, depicted in \encircle{D}, utilizes a \emph{delay thread} to precisely manage $\Delta L$ for each receive. When the progress thread receives a message, it marks it with a timestamp, $t_m$, and passes it to the delay thread. This thread queues the message, releasing it to the application only when the current time matches $t_m + \Delta L$. This enables us to precisely emulate network latency, achieving the intended delay effect.

Fig.~\ref{fig:latency-injector-setup} in the appendix provides a schematic of our latency injector and its components. We implemented our solution in MPICH and UCX. MPICH was chosen since it already has asynchronous progress threads, unlike Open MPI. UCX was favored over libfabric for its simpler API, allowing easier customization. While one can place the delay queue on the sender side, this would require capturing send requests and manipulating data buffers. This would introduce overhead from memory copies and demand extensive modifications to the MPICH source code, potentially compromising its integrity. In conclusion, our solution stands out among other methods for its portability across all UCX transports. Moreover, it allows users to conduct cluster-wide experiments without requiring special privileges or changes to existing infrastructure.

\begin{figure*}[!t]
    \centering
    \includegraphics[width=1\linewidth]{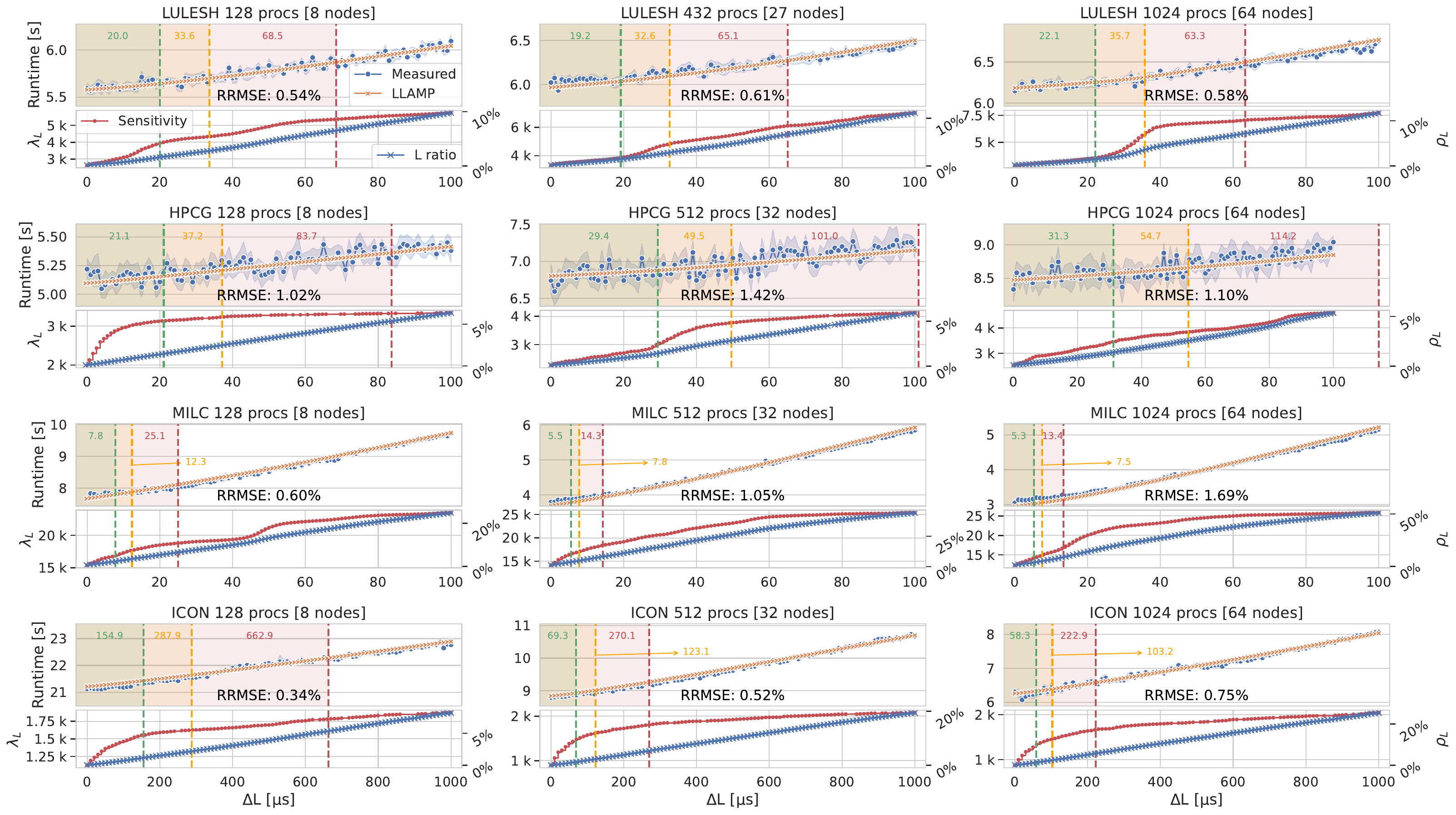}
    \cprotect\caption{Results for 4 selected applications, detailing the analysis across varying node and process counts. The upper section of each subplot compares actual runtimes against LLAMP predictions, with different colors and labels indicating \textcolor{ForestGreen}{1\%}, \textcolor{orange}{2\%}, and \textcolor{red}{5\%} $L$ tolerance. \emph{RRMSE} values highlight LLAMP's prediction accuracy relative to measured runtimes. The lower section illustrates the variation in latency sensitivity, $\lambda_L$, and the latency ratio, $\rho_L$, against changes in $\Delta L$.}
    \vspace{-1.5em}
    \label{fig:validation_results}
\end{figure*}

\subsection{Experimental Setup}

All the experiments detailed in this section were conducted on a 188-node test-bed cluster maintained by the Swiss National Supercomputing Center (CSCS). The cluster features a fat-tree topology built on 18 Mellanox SX6036 switches. Each node is powered by a 20-core Intel Xeon CPU E5-2660 v2, equipped with 32 GB DDR3 RAM and a ConnectX-3 56 Gbit/s NIC, running CentOS 7.3. We used MPICH 4.1.2, UCX 1.16.0. The stack and all applications were compiled with GCC 11.4.0. To precisely measure the network parameters critical for the LogGPS model, we employed \emph{Netgauge} 2.4.6~\cite{netguage_hoefler}. Aggregating measurements across the cluster yielded network parameters: $L = 3.0\:\mathrm{\mu s}$, $G = 0.018\:\mathrm{ns}$, and $S = 256\:\mathrm{KB}$. Observing that $o > g$ across all data sizes, we opted to omit $g$ from the analysis. To enhance the accuracy of runtime predictions, we computed average packet sizes for each application from the traces and matched the $o$ value using Netgauge's outputs. Applications were executed in a hybrid MPI+OpenMP mode in which every node runs one MPI rank and 16 OpenMP threads. For each variation in network latency, $\Delta L$, we conducted 10 runs per application, averaging the runtimes to produce the final results.

\subsection{Validation Results}
\label{sec:validation-results}

\begin{figure*}[!t]
    \centering
    \includegraphics[width=1\linewidth]{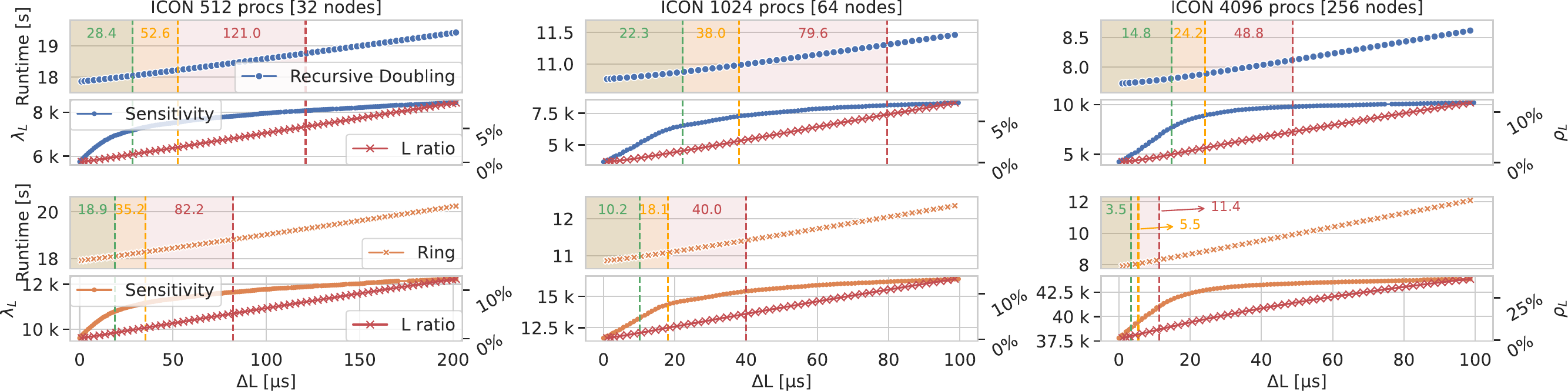}
    \cprotect\caption{Comparative analysis of ICON's network latency sensitivity and tolerance when employing recursive doubling and the ring algorithm for \code{MPI_Allreduce}. The top plots illustrate ICON's performance prediction with recursive doubling for allreduce The corresponding bottom series showcases performance outcomes utilizing the ring algorithm for the same operation.}
    \vspace{-1.5em}
    \label{fig:icon-case-study-collective}
\end{figure*}

Validation experiments were conducted on seven HPC applications across diverse domains, employing a variety of node configurations. Collecting runtime data required over 45 hours of cluster compute time. Conversely, LLAMP demonstrated its efficiency by generating equivalent results, along with additional metrics, in around 4 hours, which included time for tracing and generating LP models. Under the same condition, LogGOPSim is projected to need over 12 hours for simulation, an estimation that does not take into account LogGOPSim's inability to evaluate $\lambda_L$. Due to space limits, we showcase the outcomes from the four applications that best represent our findings in Fig.~\ref{fig:validation_results}. Detailed results and configurations for all evaluated applications are listed in Appendix~\ref{appendix:sec:validation}.

The plots demonstrate LLAMP's capability to accurately forecast the runtime of applications across various $\Delta L$, evidenced by the fact that relative root mean square errors (RRMSE)~\cite{evaluation_of_empirical_models_despotovic} is consistently under 2\%. A noticeable variation in the measured runtimes for HPCG suggests that it might be inherently more susceptible to system and network noise.
Additionally, a slight bias is observed for MILC on setups involving 32 and 64 nodes, particularly within the $0$ to $20\:\mu s$ $\Delta L$ range. This bias may stem from MILC's extensive use of persistent operations, in contrast to the \code{MPI_Send} and \code{MPI_Recv} operations utilized by Netgauge for measuring LogGP parameters. The differing overheads associated with persistent operations could contribute to this discrepancy.

Based on the definition of $\lambda_L$ (i.e., $\partial T / \partial L$), the bottom plots allow us to assess how sensitive an application is to network latency for any given $\Delta L$. The visualizations clearly mark the intervals where $\lambda_L$ increases most rapidly and where it reaches a plateau, indicating a convergence of the number of messages along the critical path toward the longest message chain in the execution graph. For example, the top right plot illustrates that LULESH, when executed on 64 nodes, has a $\lambda_L$ that is stable between $0\:\mu s$ and $20\:\mu s$. This shows that LULESH's performance remains relatively unaffected by variations in $L$ within this interval. Developers are encouraged to optimize their applications to improve communication-computation overlap, aiming for a flatter $\lambda_L$ curve.

LULESH and HPCG were evaluated under weak scaling, showing that their network latency tolerance remained relatively stable as the number of nodes increased. Interestingly, HPCG's latency tolerance improved, likely due to the optimized overlap between communication and computation. On the other hand, MILC (su3\_rmd) and ICON were subjected to strong scaling, revealing a substantial decrease in latency tolerance as the applications scaled up.  This decline is attributed to the diminishing computational workload per node in strong scaling scenarios, reducing the amount of computation that may be overlapped. 
MILC, in particular, showed a significantly lower tolerance, highlighting its great reliance on communication. 

\section{ICON Case Study}

Having validated LLAMP, we now turn our attention to applying it to a real-world application. For this purpose, we selected the \underline{Ico}sahedral \underline{N}onhydrostatic Weather and Climate Model (ICON) as our case study. ICON is a global atmospheric simulation framework designed for weather forecasting and climate studies, leveraging a novel icosahedral grid for higher precision and computational efficiency. Central to ICON is its dynamical core that solves the nonhydrostatic equations of motion. Our choice of ICON is motivated by its widespread adoption among prominent European weather services, including MeteoSwiss~\cite{icon_meteoswiss} and the German Weather Service (DWD)~\cite{icon_dwd}. Given the importance of climate studies, ICON stood out as a perfect candidate to showcase the versatile functionalities of our toolchain.

The case study was conducted on the Piz Daint supercomputer, operated by CSCS. For this study, we deployed ICON version 2.6.7, compiled with Cray MPICH 12.0.3. The LogGPS parameters measured for the cluster were $L = 1.4\:\mu s$, $G = 0.013\:\mathrm{ns}$, and $S = 256\:\mathrm{KB}$. ICON was executed in hybrid mode, utilizing 16 OpenMP threads per node. The experiments were conducted across three different scales: 32, 64, and 256 nodes. The values of $o$ for these configurations were measured as $8.5\:\mathrm{\mu s}$, $7.4\:\mathrm{\mu s}$, and $6.03\:\mathrm{\mu s}$, respectively.

\subsubsection{Impact of Collective Algorithms}

Firstly, we demonstrate LLAMP's ability to analyze how different implementations of collective algorithms impact ICON's performance. Given ICON's reliance on allreduce for data exchange in its dynamical core, we explored the effect of changing its allreduce from recursive doubling to the ring algorithm. As discussed in Section~\ref{sec:execution-graphs}, this can be achieved by altering the scheduling for allreduce in Schedgen. The results are presented in Fig.~\ref{fig:icon-case-study-collective}, where ICON was traced once per node configuration.

Our observations reveal that ICON's performance becomes increasingly sensitive to $L$ when employing the ring allreduce, a consequence of dependent sends and receives in this algorithm. This effect intensifies with the scaling of the application. Notably, at 256 nodes, ICON's $5\%\:L$ tolerance using ring allreduce stands at $11.4\:\mu s$, in contrast to its recursive doubling counterpart, which exhibits a $4\times$ latency tolerance. Furthermore, despite similar trends in their $\lambda_L$ curves across varying scales, the magnitude of $\lambda_L$ for the ring allreduce significantly exceeds that of recursive doubling, indicating a significant increase in latency sensitivity. At 256 nodes, ICON's $\rho_L$ surpasses $25\%$ for a $\Delta L = 100\:\mu s$, doubling the $\rho_L$ observed with recursive doubling. This suggests that ring allreduce should be used cautiously, especially when bandwidth and network congestion are less critical.

Through LLAMP, we highlight how software engineers can assess the impacts of collective algorithms. This approach enables a more informed decision-making process in optimizing collective communications for HPC applications.

\subsubsection{Impact of Network Topology}

LLAMP's utility extends beyond the evaluation of collective algorithms, it can also analyze various network topologies. This capability is illustrated in Fig.~\ref{fig:topology-example} in the appendix. The core concept revolves around substituting the latency of all wires with a decision variable $\l_{\mathrm{wire}}$. Given a constant switch latency, denoted as $d_{\mathrm{latency}}$, the cost to transmit a message between two nodes can be formulated as $(h + 1)\cdot l_{\mathrm{wire}} + h\cdot d_{\mathrm{latency}}$, where $h$ represents the number of hops. This allows us to model the communication cost between any two nodes as a function of $l_{\mathrm{wire}}$ and the number of hops as defined by the network topology. By setting $l_{\mathrm{wire}} \geq L$, we effectively focus on wire latency instead of the end-to-end latency between nodes. Consequently, by calculating $\partial T / \partial L$, we can determine how sensitive an application is to changes in wire latency.
This adjustment shifts the focus of analysis toward understanding the influence of the topology and the wire latency on application performance.

In this case study, we leverage LLAMP to assess the impact of two predominant network topologies, namely Fat Tree~\cite{fat_tree_alfares} and Dragonfly~\cite{dragonfly_kim}, on ICON's performance. Based on the work of Zambre et al.~\cite{breaking_band_zambre}, we set initial $l_{\mathrm{wire}}$ to $274\:\mathrm{ns}$ and $d_{\mathrm{switch}}$ to $108\:\mathrm{ns}$. For the Fat Tree topology, we chose a three-tier design where each switch has a radix of 16 ($k = 16$). The Dragonfly topology configuration selected is characterized by $g = 8$, $a = 4$, and $p = 8$. We disregard $h$ in our calculations, assuming that routing is always minimal.  Moreover, in our analysis, we assume nodes are densely packed within both topologies. For example, nodes 0 to 7 are clustered within the same pod in the Fat Tree configuration and under a single switch for the Dragonfly topology.

\begin{figure}[!htp]
    \centering
    \includegraphics[width=1\linewidth]{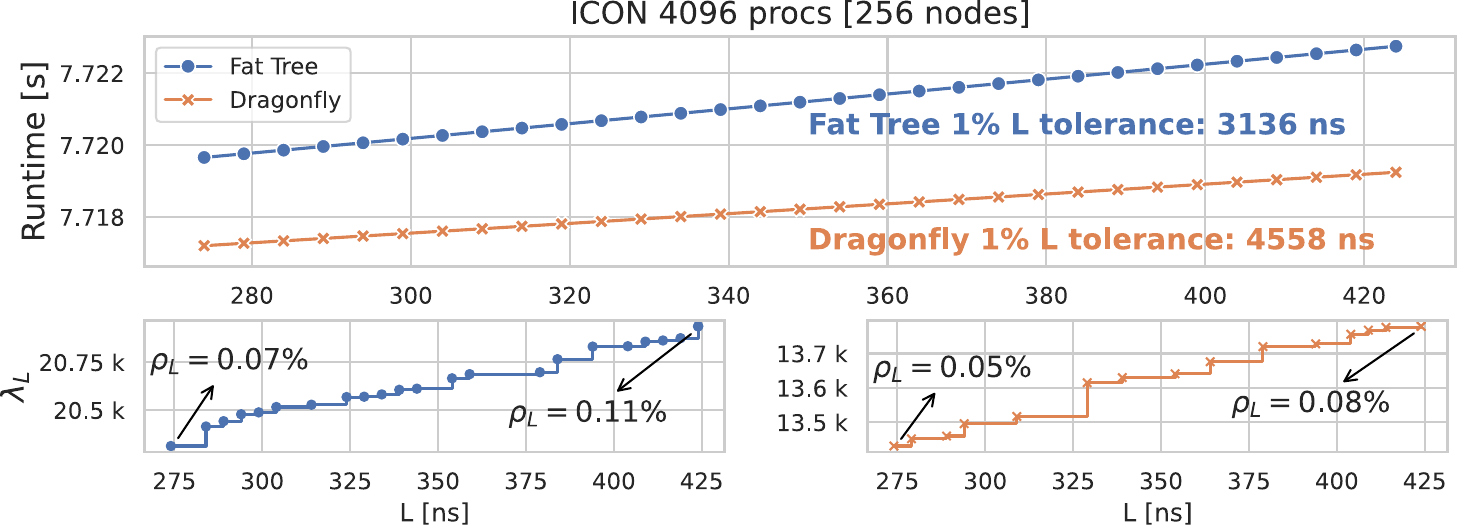}
    \caption{Comparison of the impact of the Fat Tree and Dragonfly topologies on the performance of ICON.}
    \label{fig:icon-case-study-topology}
\end{figure}

We evaluate the performance impact of these two topologies by using the execution graph of the 256-node setup (with recursive doubling allreduce) and present the result in Fig.~\ref{fig:icon-case-study-topology}. Considering the potential for a more complex FEC to increase the latency per wire by over $100\:\mathrm{ns}$ as mentioned in the introduction, we selected an interval ranging from the base latency of $274\:\mathrm{ns}$ to $424\:\mathrm{ns}$.  Our results show that ICON under Dragonfly exhibits a marginally higher network latency tolerance compared to the Fat Tree. This advantage is attributed to Dragonfly's lower average number of hops between nodes under our configurations.
The key takeaway is that despite Dragonfly's slightly superior network latency tolerance, the performance of ICON under both topologies remains largely unaffected by the anticipated latency increases from more complex FEC. This is demonstrated by the requirement for the \emph{per-link latency} to exceed $3000\:\mathrm{ns}$ before ICON's performance degrades by 1\%.

This study, while hypothetical, showcases LLAMP's adaptability in assessing the impact of network topologies. It highlights its utility for architects aiming to refine their systems for optimal performance with specific HPC applications.

\section{Related Work}

\paragraph{Trace Replay} Numerous tools exist for forecasting MPI applications' performance based on collected traces. \emph{PHANTOM}\cite{phantom_zhai, Loggpo_chen} and LogGOPSim\cite{loggopsim_hoefler} utilize communication models from the LogP family, offering fast simulation times. PSINS~\cite{psins_tikir} presents a choice among three communication models of varying complexity, with the PMaC model~\cite{pmac_snavely} standing out for its accuracy. Hermanns et al.\cite{verifying_causality_hermanns} aim at reenacting communication operations from traces to identify load imbalances. SMPI\cite{smpi_degomme} and the work of Desprez et al.~\cite{assessing_the_performance_desprez}, rely on SimGrid~\cite{simgrid_casanova}, a discrete event simulator, for more precise performance predictions during online simulations. Kenny et al.\cite{the_pitfalls_of_provisioning_kenny} incorporate SST\cite{sst_rodrigues} and PISCES, a packet-level model, for architectural simulations, employing Bayesian inference for simulator validation. Eyerman et al.~\cite{accurate_and_scalable_eyerman} propose integrating profiling, node simulation, and high-level network simulation with SST. Despite their capabilities, these simulators often rely on complex discrete event and packet-level simulators that are usually time-consuming to execute. One significant limitation they share, when compared to LLAMP, is the necessity for extensive parameter sweeps to ascertain each application's network latency tolerance.

\paragraph{Critical Path Analysis}
Graph-based analysis of MPI applications is a well-established approach. Schulz~\cite{extracting_critical_path_schulz} and Böhme et al.\cite{scalable_critical_path_bohme} have contributed to identifying wait-states and load imbalances by employing critical path analysis and trace replay. Chen et al.\cite{critical_path_candidates_chen} further refined this approach by introducing the concept of critical-path candidates, identifying potential critical paths through profiling-based instruction and communication counts. Schmitt et al.~\cite{scalable_critical_path_hybrid_schmitt} expanded these mostly rely on traditional graph analysis, LLAMP introduces a completely new and unique perspective on viewing graphs and extracting information from the critical path.

\paragraph{Network Latency Sensitivity and Tolerance}
Numerous studies have investigated how network latency affects application performance. Efforts by Underwood et al.~\cite{measuring_network_latency_underwood}, Gao et al.\cite{network_requirements_gao}, Patke et al.~\cite{evaluating_hardware_memory_disaggregation_patke}, Link Gradients~\cite{link_gradients_chen}, and Richar Paul Martin's thesis~\cite{a_systematic_characterization_martin} have predominantly involved manually injecting latency to networks and conducting extensive trials with varied latency levels to identify patterns. Kerbyson et al.~\cite{a_look_kerbyson} utilized performance models of three specific applications for similar analyses. While insightful, it demands considerable effort and deep application knowledge, limiting their generality. Our work stands out by mathematically defining network latency sensitivity/tolerance and offering a scalable, analytical method to compute these metrics across a broad spectrum of MPI applications.

\section{Extensions and Discussion}

Beyond the capabilities we have shown, LLAMP holds extensive potential for a broad spectrum of analyses, offering users a flexible framework to explore a variety of performance metrics. For example, LLAMP's versatility extends to examining the sensitivity and tolerance of additional critical LogGPS model parameters, like bandwidth $G$. Its support for \emph{heterogeneous network models}, such as HLogGP, unlocks the potential to assess pairwise network latency sensitivities across MPI ranks. Leveraging this, we devised a new process placement algorithm to optimize the mapping of MPI ranks onto physical processors. These functionalities, detailed in Appendices~\ref{appendix:sec:heterogenenous-loggp} and \ref{appendix:sec:rank-placement}, exemplify LLAMP's versatility and how it empowers users to tailor the toolchain to their specific needs.

There are aspects of LLAMP that can still be improved. Firstly, the trace-based nature of our analysis means that LLAMP relies on a static snapshot of application behavior to produce the performance metrics, which may not capture the variability of live execution. As exemplified by the works of Nikitenko et al.~\cite{influence_of_noisy_environments_nikitenko} and Hoefler et al.~\cite{characterizing_the_influence_hoefler}, an application's performance during tracing might be influenced by system noise and network congestion. To overcome this, we can incorporate a more detailed model, such as LogGOPSC~\cite{loggopsc_yan}, to account for network contention or statistics to account for the variability of network behavior and system noise, allowing LLAMP to predict a range of outcomes and their probabilities.

\begin{figure}[!htp]
    \centering
    \includegraphics[width=1\linewidth]{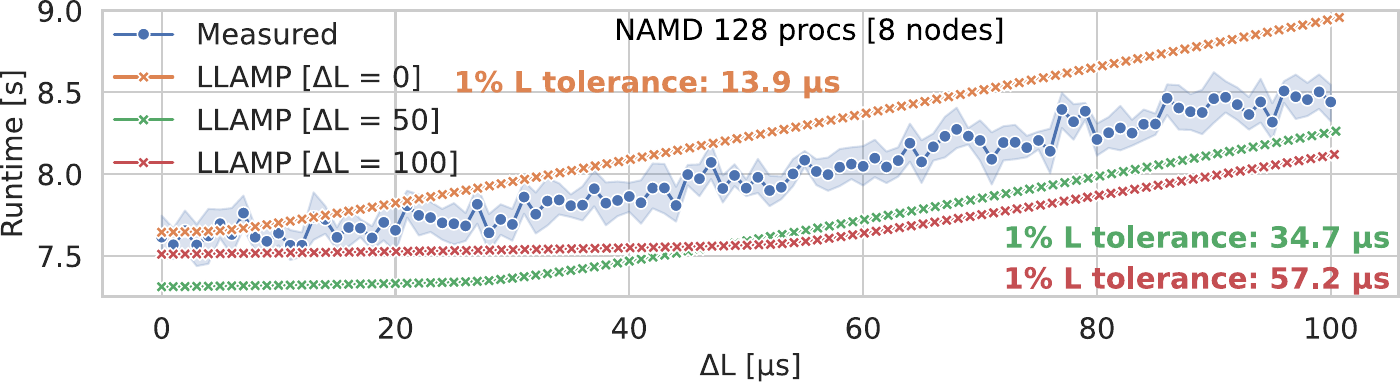}
    \caption{Runtime of NAMD, a molecular dynamics simulation built on charm++, showing measured vs. predicted runtimes. The legend indicates $\Delta L$ at which the traces were recorded.}
    \label{fig:namd-discussion}
\end{figure}

Currently, LLAMP's is not fully equipped to handle the intricacies of dynamically scheduled operations seen in frameworks like \emph{charm++}\cite{charm_kale}, which dynamically adjust based on runtime conditions. However, LLAMP still manages to reveal the inherent latency tolerance within these frameworks. As illustrated in Fig.~\ref{fig:namd-discussion}, LLAMP demonstrates how charm++ proactively adjusts its communication schedule to enhance resilience to increased network latency. In the future, we plan to explore a general strategy that can analyze and quantify network latency sensitivity and tolerance across a variety of parallel programming models. This includes not only those that use dynamic scheduling but also models that diverge from the MPI standard, such as Legion~\cite{legion_bauer}.

\section{Conclusion}

In this work, we introduce LLAMP, an innovative toolchain that integrates the LogGPS model with linear programming to analyze network latency sensitivity and tolerance in HPC applications. By uncovering the intricate connection between the critical paths in a program's execution graphs and LP, we provide a new method for efficiently gathering performance metrics. LLAMP's flexibility allows users to utilize metrics such as $\lambda_L$ and $\rho_L$ in ways tailored to their specific needs.

Moreover, the development of a specialized latency injector that accurately emulates flow-level latency enables us to validate LLAMP's predictions across a variety of HPC applications, demonstrating an error margin of less than 2\%. To illustrate LLAMP's practicality, we conducted a case study with ICON, a leading climate modeling tool in Europe, demonstrating its utility in addressing the needs of both software engineers and network architects.

\section*{Acknowledgements}

The authors would like to thank Andrei Ivanov and Lukas Gianinazzi for their helpful discussions. We also wish to thank the Swiss National Supercomputing Centre (CSCS) for their support of this work.
This work was financially supported by the 2023 Global Research Outreach Program of the Samsung Advanced Institute of Technology (SAIT) and Samsung R\&D Center America, Silicon Valley (SRA-SV) under the supervision of the System Architecture Laboratory (SAL). 
This work received support from the PASC project DaCeMI: Harnessing future hardware using Data-Centric ML Integration.
This project has received funding from the European High-Performance Computing Joint Undertaking (JU) under grant agreement No 955776 and a donation from Intel.

\bibliographystyle{ieeetr}
\bibliography{references}

\begin{thebibliography}{10}

\bibitem{meta_ai_center_wagner}
K.~Wagner, ``Meta is building new \$800 million ai-focused data center in
  indiana,'' {\em Bloomberg}.

\bibitem{cscs_alps}
CSCS, ``Cscs, hewlett packard enterprise and nvidia announce world’s most
  powerful ai-capable supercomputer,'' {\em Swiss National Supercomputing
  Center}.

\bibitem{cloud_computing_close_gap_guidi}
G.~Guidi, M.~Ellis, A.~Bulu\c{c}, K.~Yelick, and D.~Culler, ``10 years later:
  Cloud computing is closing the performance gap,'' in {\em Companion of the
  ACM/SPEC International Conference on Performance Engineering}, ICPE '21, (New
  York, NY, USA), p.~41–48, Association for Computing Machinery, 2021.

\bibitem{cost_effective_hpc_carlyle}
A.~G. Carlyle, S.~L. Harrell, and P.~M. Smith, ``Cost-effective hpc: The
  community or the cloud?,'' in {\em 2010 IEEE Second International Conference
  on Cloud Computing Technology and Science}, pp.~169--176, 2010.

\bibitem{noise_in_the_clouds_de_sensi}
D.~De~Sensi, T.~De~Matteis, K.~Taranov, S.~Di~Girolamo, T.~Rahn, and
  T.~Hoefler, ``Noise in the clouds: Influence of network performance
  variability on application scalability,'' {\em Proc. ACM Meas. Anal. Comput.
  Syst.}, vol.~6, dec 2022.

\bibitem{evaluation_of_hpc_applications_gupta}
A.~Gupta and D.~Milojicic, ``Evaluation of hpc applications on cloud,'' in {\em
  2011 Sixth Open Cirrus Summit}, pp.~22--26, 2011.

\bibitem{a_comparative_study_marathe}
A.~Marathe, R.~Harris, D.~K. Lowenthal, B.~R. de~Supinski, B.~Rountree,
  M.~Schulz, and X.~Yuan, ``A comparative study of high-performance computing
  on the cloud,'' in {\em Proceedings of the 22nd International Symposium on
  High-Performance Parallel and Distributed Computing}, HPDC '13, (New York,
  NY, USA), p.~239–250, Association for Computing Machinery, 2013.

\bibitem{performance_analysis_of_hpc_exposito}
R.~R. Exp{\'o}sito, G.~L. Taboada, S.~Ramos, J.~Touri{\~n}o, and R.~Doallo,
  ``Performance analysis of hpc applications in the cloud,'' {\em Future
  Generation Computer Systems}, vol.~29, no.~1, pp.~218--229, 2013.
\newblock Including Special section: AIRCC-NetCoM 2009 and Special section:
  Clouds and Service-Oriented Architectures.

\bibitem{running_hpc_tomic}
D.~Tomić, Z.~Car, and D.~Ogrizović, ``Running hpc applications on many
  million cores cloud,'' in {\em 2017 40th International Convention on
  Information and Communication Technology, Electronics and Microelectronics
  (MIPRO)}, pp.~209--214, 2017.

\bibitem{deep_learning_training_naumov}
M.~Naumov, J.~Kim, D.~Mudigere, S.~Sridharan, X.~Wang, W.~Zhao, S.~Yilmaz,
  C.~Kim, H.~Yuen, M.~Ozdal, K.~Nair, I.~Gao, B.-Y. Su, J.~Yang, and
  M.~Smelyanskiy, ``Deep learning training in facebook data centers: Design of
  scale-up and scale-out systems,'' 2020.

\bibitem{varuna_athlur}
S.~Athlur, N.~Saran, M.~Sivathanu, R.~Ramjee, and N.~Kwatra, ``Varuna:
  scalable, low-cost training of massive deep learning models,'' in {\em
  Proceedings of the Seventeenth European Conference on Computer Systems},
  EuroSys '22, (New York, NY, USA), p.~472–487, Association for Computing
  Machinery, 2022.

\bibitem{architectural_requirements_ibrahim}
K.~Z. Ibrahim, T.~Nguyen, H.~A. Nam, W.~Bhimji, S.~Farrell, L.~Oliker,
  M.~Rowan, N.~J. Wright, and S.~Williams, ``Architectural requirements for
  deep learning workloads in hpc environments,'' in {\em 2021 International
  Workshop on Performance Modeling, Benchmarking and Simulation of High
  Performance Computer Systems (PMBS)}, pp.~7--17, 2021.

\bibitem{using_ml_to_model_barrachina}
S.~Barrachina, A.~Castell{\'o}, M.~Catal{\'a}n, M.~F. Dolz, and J.~I. Mestre,
  ``Using machine learning to model the training scalability of convolutional
  neural networks on clusters of gpus,'' {\em Computing}, vol.~105, no.~5,
  pp.~915--934, 2023.

\bibitem{datacenter_ethernet_hoefler}
T.~Hoefler, D.~Roweth, K.~Underwood, R.~Alverson, M.~Griswold, V.~Tabatabaee,
  M.~Kalkunte, S.~Anubolu, S.~Shen, M.~McLaren, A.~Kabbani, and S.~Scott,
  ``Data center ethernet and remote direct memory access: Issues at
  hyperscale,'' {\em Computer}, vol.~56, no.~7, pp.~67--77, 2023.

\bibitem{ieee_dambrosia}
J.~D’Ambrosia, ``Ieee p802.3df™ defines architecture holistically to
  achieve 800 gb/s and 1.6 tb/s ethernet,'' {\em IEEE Standards Association},
  2022.

\bibitem{100_gbs_ethernet_fec_liu}
C.~Liu, ``{100+ Gb/s Ethernet Forward Error Correction (FEC) Analysis},'' tech.
  rep., 2019.

\bibitem{fec_for_400g_bates}
S.~Bates, ``Forward error correction for 400g - ieee 802,'' 2013.

\bibitem{hammingmesh_hoefler}
T.~Hoefler, T.~Bonato, D.~De~Sensi, S.~Di~Girolamo, S.~Li, M.~Heddes, J.~Belk,
  D.~Goel, M.~Castro, and S.~Scott, ``Hammingmesh: A network topology for
  large-scale deep learning,'' in {\em SC22: International Conference for High
  Performance Computing, Networking, Storage and Analysis}, pp.~1--18, 2022.

\bibitem{scalable_deep_learning_mayer}
R.~Mayer and H.-A. Jacobsen, ``Scalable deep learning on distributed
  infrastructures: Challenges, techniques, and tools,'' {\em ACM Comput.
  Surv.}, vol.~53, feb 2020.

\bibitem{milc_bernard}
C.~Bernard, M.~C. Ogilvie, T.~A. Degrand, C.~E. Detar, S.~A. Gottlieb,
  A.~Krasnitz, R.~Sugar, and D.~Toussaint, ``Studying quarks and gluons on mimd
  parallel computers,'' {\em Int. J. High Perform. Comput. Appl.}, vol.~5,
  p.~61–70, dec 1991.

\bibitem{lulesh_2_karlin}
I.~Karlin, J.~Keasler, and R.~Neely, ``Lulesh 2.0 updates and changes,'' Tech.
  Rep. LLNL-TR-641973, August 2013.

\bibitem{icon_zangl}
G.~Z{\"a}ngl, D.~Reinert, P.~R{\'\i}podas, and M.~Baldauf, ``The icon
  (icosahedral non-hydrostatic) modelling framework of dwd and mpi-m:
  Description of the non-hydrostatic dynamical core,'' {\em Quarterly Journal
  of the Royal Meteorological Society}, vol.~141, no.~687, pp.~563--579, 2015.

\bibitem{a_look_kerbyson}
D.~Kerbyson, ``A look at application performance sensitivity to the bandwidth
  and latency of infiniband networks,'' in {\em Proceedings 20th IEEE
  International Parallel \& Distributed Processing Symposium}, pp.~7 pp.--,
  2006.

\bibitem{performance_modeling_milc_bauer}
G.~Bauer, S.~Gottlieb, and T.~Hoefler, ``Performance modeling and comparative
  analysis of the milc lattice qcd application su3\_rmd,'' in {\em 2012 12th
  IEEE/ACM International Symposium on Cluster, Cloud and Grid Computing (ccgrid
  2012)}, pp.~652--659, 2012.

\bibitem{evaluating_hardware_memory_disaggregation_patke}
A.~Patke, H.~Qiu, S.~Jha, S.~Venugopal, M.~Gazzetti, C.~Pinto, Z.~Kalbarczyk,
  and R.~Iyer, ``Evaluating hardware memory disaggregation under delay and
  contention,'' in {\em 2022 IEEE International Parallel and Distributed
  Processing Symposium Workshops (IPDPSW)}, pp.~1221--1227, 2022.

\bibitem{characterizing_application_rosenthal}
E.~Rosenthal and E.~Leon, ``Characterizing application sensitivity to network
  performance,'' 11 2014.

\bibitem{analyzing_cost_performance_bhatele}
A.~Bhatele, N.~Jain, M.~Mubarak, and T.~Gamblin, ``Analyzing cost-performance
  tradeoffs of hpc network designs under different constraints using
  simulations,'' in {\em Proceedings of the 2019 ACM SIGSIM Conference on
  Principles of Advanced Discrete Simulation}, SIGSIM-PADS '19, (New York, NY,
  USA), p.~1–12, Association for Computing Machinery, 2019.

\bibitem{towards_million_server_besta}
M.~Besta, M.~Schneider, S.~D. Girolamo, A.~Singla, and T.~Hoefler, ``Towards
  million-server network simulations on just a laptop,'' {\em CoRR},
  vol.~abs/2105.12663, 2021.

\bibitem{simulation_based_xu}
G.~Xu, H.~Ibeid, X.~Jiang, V.~Svilan, and Z.~Bian, ``Simulation-based
  performance prediction of hpc applications: A case study of hpl,'' in {\em
  2020 IEEE/ACM International Workshop on HPC User Support Tools (HUST) and
  Workshop on Programming and Performance Visualization Tools (ProTools)},
  pp.~81--88, 2020.

\bibitem{enabling_parallel_mubarak}
M.~Mubarak, C.~D. Carothers, R.~B. Ross, and P.~Carns, ``Enabling parallel
  simulation of large-scale hpc network systems,'' {\em IEEE Transactions on
  Parallel and Distributed Systems}, vol.~28, no.~1, pp.~87--100, 2017.

\bibitem{loggopsim_hoefler}
T.~Hoefler, T.~Schneider, and A.~Lumsdaine, ``Loggopsim: Simulating large-scale
  applications in the loggops model,'' in {\em Proceedings of the 19th ACM
  International Symposium on High Performance Distributed Computing}, HPDC '10,
  (New York, NY, USA), p.~597–604, Association for Computing Machinery, 2010.

\bibitem{extracting_critical_path_schulz}
M.~Schulz, ``Extracting critical path graphs from mpi applications,'' in {\em
  2005 IEEE International Conference on Cluster Computing}, pp.~1--10, 2005.

\bibitem{scalable_critical_path_bohme}
D.~Böhme, F.~Wolf, B.~R. de~Supinski, M.~Schulz, and M.~Geimer, ``Scalable
  critical-path based performance analysis,'' in {\em 2012 IEEE 26th
  International Parallel and Distributed Processing Symposium}, pp.~1330--1340,
  2012.

\bibitem{investigating_dependency_graph_pereira}
R.~Pereira, A.~Roussel, P.~Carribault, and T.~Gautier, ``Investigating
  dependency graph discovery impact on task-based mpi+openmp applications
  performances,'' in {\em Proceedings of the 52nd International Conference on
  Parallel Processing}, ICPP '23, (New York, NY, USA), p.~163–172,
  Association for Computing Machinery, 2023.

\bibitem{characterizing_the_influence_hoefler}
T.~Hoefler, T.~Schneider, and A.~Lumsdaine, ``Characterizing the influence of
  system noise on large-scale applications by simulation,'' in {\em SC '10:
  Proceedings of the 2010 ACM/IEEE International Conference for High
  Performance Computing, Networking, Storage and Analysis}, pp.~1--11, 2010.

\bibitem{using_simulation_levy}
S.~Levy, B.~Topp, K.~B. Ferreira, D.~Arnold, T.~Hoefler, and P.~Widener,
  ``Using simulation to evaluate the performance of resilience strategies at
  scale,'' in {\em High Performance Computing Systems. Performance Modeling,
  Benchmarking and Simulation} (S.~A. Jarvis, S.~A. Wright, and S.~D. Hammond,
  eds.), (Cham), pp.~91--114, Springer International Publishing, 2014.

\bibitem{what_if_mpi_collective_riesen}
R.~Riesen, C.~Vaughan, and T.~Hoefler, ``What if mpi collective operations were
  instantaneous?,'' {\em Cray User Group (CUG)}, 2006.

\bibitem{group_operation_assembly_language_hoefler}
T.~Hoefler, C.~Siebert, and A.~Lumsdaine, ``Group operation assembly language -
  a flexible way to express collective communication,'' in {\em 2009
  International Conference on Parallel Processing}, pp.~574--581, 2009.

\bibitem{logp_culler}
D.~Culler, R.~Karp, D.~Patterson, A.~Sahay, K.~E. Schauser, E.~Santos,
  R.~Subramonian, and T.~von Eicken, ``Logp: towards a realistic model of
  parallel computation,'' {\em SIGPLAN Not.}, vol.~28, p.~1–12, jul 1993.

\bibitem{a_survey_rico_gallego}
J.~A. Rico-Gallego, J.~C. D\'{\i}az-Mart\'{\i}n, R.~R. Manumachu, and A.~L.
  Lastovetsky, ``A survey of communication performance models for
  high-performance computing,'' {\em ACM Comput. Surv.}, vol.~51, jan 2019.

\bibitem{loggp_alexandrov}
A.~Alexandrov, M.~F. Ionescu, K.~E. Schauser, and C.~Scheiman, ``Loggp:
  incorporating long messages into the logp model—one step closer towards a
  realistic model for parallel computation,'' in {\em Proceedings of the
  Seventh Annual ACM Symposium on Parallel Algorithms and Architectures}, SPAA
  '95, (New York, NY, USA), p.~95–105, Association for Computing Machinery,
  1995.

\bibitem{sensitivity_analysis_borgonovo}
E.~Borgonovo and E.~Plischke, ``Sensitivity analysis: A review of recent
  advances,'' {\em European Journal of Operational Research}, vol.~248, no.~3,
  pp.~869--887, 2016.

\bibitem{sensitivity_analysis_saltelli}
A.~Saltelli, S.~Tarantola, F.~Campolongo, and M.~Ratto, {\em Sensitivity
  Analysis in Practice: A Guide to Assessing Scientific Models}.
\newblock Wiley, 2004.

\bibitem{sensitivity_analysis_of_model_output_borgonovo}
E.~Borgonovo, ``Sensitivity analysis of model output with input constraints: A
  generalized rationale for local methods,'' {\em Risk Analysis}, vol.~28,
  no.~3, pp.~667--680, 2008.

\bibitem{sensitivity_analysis_of_environmental_models_pianosi}
F.~Pianosi, K.~Beven, J.~Freer, J.~W. Hall, J.~Rougier, D.~B. Stephenson, and
  T.~Wagener, ``Sensitivity analysis of environmental models: A systematic
  review with practical workflow,'' {\em Environmental Modelling \& Software},
  vol.~79, pp.~214--232, 2016.

\bibitem{critical_path_candidates_chen}
J.~Chen and R.~M. Clapp, ``Critical-path candidates: scalable performance
  modeling for mpi workloads,'' in {\em 2015 IEEE International Symposium on
  Performance Analysis of Systems and Software (ISPASS)}, pp.~1--10, 2015.

\bibitem{linear_programming_britannica}
{The Editors of Encyclopaedia Britannica}, ``linear programming.''
  \url{https://www.britannica.com/science/linear-programming-mathematics}, Feb
  2024.
\newblock Accessed 25 March 2024.

\bibitem{lp_algorithms}
{\em Linear Programming Algorithms}, pp.~71--97.
\newblock Berlin, Heidelberg: Springer Berlin Heidelberg, 2008.

\bibitem{how_good_is_the_simplex_algorithm_klee}
V.~Klee and G.~J. Minty, ``How good is the simplex algorithm,'' 1970.

\bibitem{measuring_network_latency_underwood}
R.~Underwood, J.~Anderson, and A.~Apon, ``Measuring network latency variation
  impacts to high performance computing application performance,'' in {\em
  Proceedings of the 2018 ACM/SPEC International Conference on Performance
  Engineering}, ICPE '18, (New York, NY, USA), p.~68–79, Association for
  Computing Machinery, 2018.

\bibitem{gurobi}
{Gurobi Optimization, LLC}, ``{Gurobi Optimizer Reference Manual},'' 2023.

\bibitem{advanced_gurobi_algorithms_wunderling}
{Wunderling, Roland}, ``{Advanced Gurobi Algorithms},'' 2022.

\bibitem{presolving_in_lp_andersen}
E.~D. Andersen and K.~D. Andersen, ``Presolving in linear programming,'' {\em
  Mathematical Programming}, vol.~71, no.~2, pp.~221--245, 1995.

\bibitem{applied_mathematical_programming_bradley}
S.~Bradley, A.~Hax, and T.~Magnanti, {\em Applied Mathematical Programming}.
\newblock Addison-Wesley Publishing Company, 1977.

\bibitem{netguage_hoefler}
T.~Hoefler, T.~Mehlan, A.~Lumsdaine, and W.~Rehm, ``{Netgauge: A Network
  Performance Measurement Framework},'' in {\em Proceedings of High Performance
  Computing and Communications, HPCC'07}, vol.~4782, pp.~659--671, Springer,
  Sep. 2007.

\bibitem{evaluation_of_empirical_models_despotovic}
M.~Despotovic, V.~Nedic, D.~Despotovic, and S.~Cvetanovic, ``Evaluation of
  empirical models for predicting monthly mean horizontal diffuse solar
  radiation,'' {\em Renewable and Sustainable Energy Reviews}, vol.~56,
  pp.~246--260, 2016.

\bibitem{icon_meteoswiss}
MeteoSwiss, ``Icon-22,'' 2019.

\bibitem{icon_dwd}
DWD, ``Icon (icosahedral nonhydrostatic) model.''

\bibitem{fat_tree_alfares}
M.~Al-Fares, A.~Loukissas, and A.~Vahdat, ``A scalable, commodity data center
  network architecture,'' in {\em Proceedings of the ACM SIGCOMM 2008
  Conference on Data Communication}, SIGCOMM '08, (New York, NY, USA),
  p.~63–74, Association for Computing Machinery, 2008.

\bibitem{dragonfly_kim}
J.~Kim, W.~J. Dally, S.~Scott, and D.~Abts, ``Technology-driven,
  highly-scalable dragonfly topology,'' in {\em 2008 International Symposium on
  Computer Architecture}, pp.~77--88, 2008.

\bibitem{breaking_band_zambre}
M.~Al-Fares, A.~Loukissas, and A.~Vahdat, ``A scalable, commodity data center
  network architecture,'' in {\em Proceedings of the ACM SIGCOMM 2008
  Conference on Data Communication}, SIGCOMM '08, (New York, NY, USA),
  p.~63–74, Association for Computing Machinery, 2008.

\bibitem{phantom_zhai}
J.~Zhai, W.~Chen, and W.~Zheng, ``Phantom: predicting performance of parallel
  applications on large-scale parallel machines using a single node,'' in {\em
  Proceedings of the 15th ACM SIGPLAN Symposium on Principles and Practice of
  Parallel Programming}, PPoPP '10, (New York, NY, USA), p.~305–314,
  Association for Computing Machinery, 2010.

\bibitem{Loggpo_chen}
W.~Chen, J.~Zhai, J.~Zhang, and W.~Zheng, ``Loggpo: An accurate communication
  model for performance prediction of mpi programs,'' {\em Science in China
  Series F: Information Sciences}, vol.~52, no.~10, pp.~1785--1791, 2009.

\bibitem{psins_tikir}
M.~M. Tikir, M.~A. Laurenzano, L.~Carrington, and A.~Snavely, ``Psins: An open
  source event tracer and execution simulator for mpi applications,'' in {\em
  Euro-Par 2009 Parallel Processing} (H.~Sips, D.~Epema, and H.-X. Lin, eds.),
  (Berlin, Heidelberg), pp.~135--148, Springer Berlin Heidelberg, 2009.

\bibitem{pmac_snavely}
A.~Snavely, L.~Carrington, N.~Wolter, J.~Labarta, R.~Badia, and A.~Purkayastha,
  ``A framework for performance modeling and prediction,'' in {\em SC '02:
  Proceedings of the 2002 ACM/IEEE Conference on Supercomputing}, pp.~21--21,
  2002.

\bibitem{verifying_causality_hermanns}
M.-A. Hermanns, M.~Geimer, F.~Wolf, and B.~J. Wylie, ``Verifying causality
  between distant performance phenomena in large-scale mpi applications,'' in
  {\em 2009 17th Euromicro International Conference on Parallel, Distributed
  and Network-based Processing}, pp.~78--84, 2009.

\bibitem{smpi_degomme}
A.~Degomme, A.~Legrand, G.~S. Markomanolis, M.~Quinson, M.~Stillwell, and
  F.~Suter, ``Simulating mpi applications: The smpi approach,'' {\em IEEE
  Transactions on Parallel and Distributed Systems}, vol.~28, no.~8,
  pp.~2387--2400, 2017.

\bibitem{assessing_the_performance_desprez}
F.~Desprez, G.~S. Markomanolis, M.~Quinson, and F.~Suter, ``Assessing the
  performance of mpi applications through time-independent trace replay,'' in
  {\em 2011 40th International Conference on Parallel Processing Workshops},
  pp.~467--476, 2011.

\bibitem{simgrid_casanova}
H.~Casanova, A.~Giersch, A.~Legrand, M.~Quinson, and F.~Suter, ``Versatile,
  scalable, and accurate simulation of distributed applications and
  platforms,'' {\em Journal of Parallel and Distributed Computing}, vol.~74,
  pp.~2899--2917, June 2014.

\bibitem{the_pitfalls_of_provisioning_kenny}
J.~P. Kenny, K.~Sargsyan, S.~Knight, G.~Michelogiannakis, and J.~J. Wilke,
  ``The pitfalls of provisioning exascale networks: A trace replay analysis for
  understanding communication performance,'' in {\em High Performance
  Computing} (R.~Yokota, M.~Weiland, D.~Keyes, and C.~Trinitis, eds.), (Cham),
  pp.~269--288, Springer International Publishing, 2018.

\bibitem{sst_rodrigues}
A.~F. Rodrigues, K.~S. Hemmert, B.~W. Barrett, C.~Kersey, R.~Oldfield,
  M.~Weston, R.~Risen, J.~Cook, P.~Rosenfeld, E.~Cooper-Balis, and B.~Jacob,
  ``The structural simulation toolkit,'' {\em SIGMETRICS Perform. Eval. Rev.},
  vol.~38, p.~37–42, mar 2011.

\bibitem{accurate_and_scalable_eyerman}
S.~Eyerman, W.~Heirman, K.~D. Bois, and I.~Hur, ``Accurate and scalable
  many-node simulation,'' 2024.

\bibitem{scalable_critical_path_hybrid_schmitt}
F.~Schmitt, R.~Dietrich, and G.~Juckeland, ``Scalable critical path analysis
  for hybrid mpi-cuda applications,'' in {\em 2014 IEEE International Parallel
  \& Distributed Processing Symposium Workshops}, pp.~908--915, 2014.

\bibitem{network_requirements_gao}
P.~X. Gao, A.~Narayan, S.~Karandikar, J.~Carreira, S.~Han, R.~Agarwal,
  S.~Ratnasamy, and S.~Shenker, ``Network requirements for resource
  disaggregation,'' in {\em Proceedings of the 12th USENIX Conference on
  Operating Systems Design and Implementation}, OSDI'16, (USA), p.~249–264,
  USENIX Association, 2016.

\bibitem{link_gradients_chen}
S.~Chen, K.~R. Joshi, M.~A. Hiltunen, W.~H. Sanders, and R.~D. Schlichting,
  ``Link gradients: Predicting the impact of network latency on multitier
  applications,'' in {\em IEEE INFOCOM 2009}, pp.~2258--2266, 2009.

\bibitem{a_systematic_characterization_martin}
R.~P. Martin, {\em A Systematic Characterization of Application Sensitivity to
  Network Performance}.
\newblock PhD thesis, EECS Department, University of California, Berkeley,
  1999.

\bibitem{influence_of_noisy_environments_nikitenko}
D.~A. Nikitenko, F.~Wolf, B.~Mohr, T.~Hoefler, K.~S. Stefanov, V.~V. Voevodin,
  A.~S. Antonov, and A.~Calotoiu, ``Influence of noisy environments on behavior
  of hpc applications,'' {\em Lobachevskii Journal of Mathematics}, vol.~42,
  no.~7, pp.~1560--1570, 2021.

\bibitem{loggopsc_yan}
Y.~Baicheng, Z.~Yi, X.~Limin, H.~Jiantong, and W.~Zhaokai, ``Loggopsc: A
  parallel computation model extending network contention into loggops,'' in
  {\em 2019 IEEE International Conference on Cluster Computing (CLUSTER)},
  pp.~1--2, 2019.

\bibitem{charm_kale}
L.~V. Kale and S.~Krishnan, ``Charm++: a portable concurrent object oriented
  system based on c++,'' {\em SIGPLAN Not.}, vol.~28, p.~91–108, oct 1993.

\bibitem{legion_bauer}
M.~Bauer, S.~Treichler, E.~Slaughter, and A.~Aiken, ``Legion: Expressing
  locality and independence with logical regions,'' in {\em SC '12: Proceedings
  of the International Conference on High Performance Computing, Networking,
  Storage and Analysis}, pp.~1--11, 2012.

\bibitem{loggps_ino}
F.~Ino, N.~Fujimoto, and K.~Hagihara, ``Loggps: a parallel computational model
  for synchronization analysis,'' in {\em Proceedings of the Eighth ACM SIGPLAN
  Symposium on Principles and Practices of Parallel Programming}, PPoPP '01,
  (New York, NY, USA), p.~133–142, Association for Computing Machinery, 2001.

\bibitem{npb_bailey}
D.~H. Bailey, {\em NAS Parallel Benchmarks}, pp.~1254--1259.
\newblock Boston, MA: Springer US, 2011.

\bibitem{lammps_thompson}
A.~P. Thompson, H.~M. Aktulga, R.~Berger, D.~S. Bolintineanu, W.~M. Brown,
  P.~S. Crozier, P.~J. in~'t Veld, A.~Kohlmeyer, S.~G. Moore, T.~D. Nguyen,
  R.~Shan, M.~J. Stevens, J.~Tranchida, C.~Trott, and S.~J. Plimpton,
  ``{LAMMPS} - a flexible simulation tool for particle-based materials modeling
  at the atomic, meso, and continuum scales,'' {\em Comp. Phys. Comm.},
  vol.~271, p.~108171, 2022.

\bibitem{toward_a_new_metric_heroux}
M.~A. Heroux and J.~Dongarra, ``Toward a new metric for ranking high
  performance computing systems.,'' 6 2013.

\bibitem{openmx_boker}
S.~M. Boker, M.~C. Neale, H.~H. Maes, M.~Spiegel, T.~R. Brick, R.~Estabrook,
  T.~C. Bates, R.~J. Gore, M.~D. Hunter, J.~N. Pritikin, M.~Zahery, and R.~M.
  Kirkpatrick, {\em OpenMx: Extended Structural Equation Modelling}, 2023.
\newblock R package version 2.21.11.

\bibitem{cloverleaf_mallinson}
A.~Mallinson, D.~A. Beckingsale, W.~Gaudin, J.~Herdman, J.~Levesque, and S.~A.
  Jarvis, ``Cloverleaf: Preparing hydrodynamics codes for exascale,'' {\em The
  Cray User Group}, vol.~2013, 2013.

\bibitem{hloggp_bosque}
J.~Bosque and L.~Perez, ``Hloggp: a new parallel computational model for
  heterogeneous clusters,'' in {\em IEEE International Symposium on Cluster
  Computing and the Grid, 2004. CCGrid 2004.}, pp.~403--410, 2004.

\bibitem{process_mapping_zhang}
J.~Zhang, J.~Zhai, W.~Chen, and W.~Zheng, ``Process mapping for mpi collective
  communications,'' in {\em Euro-Par 2009 Parallel Processing} (H.~Sips,
  D.~Epema, and H.-X. Lin, eds.), (Berlin, Heidelberg), pp.~81--92, Springer
  Berlin Heidelberg, 2009.

\bibitem{mpipp_chen}
H.~Chen, W.~Chen, J.~Huang, B.~Robert, and H.~Kuhn, ``Mpipp: An automatic
  profile-guided parallel process placement toolset for smp clusters and
  multiclusters,'' in {\em Proceedings of the 20th Annual International
  Conference on Supercomputing}, ICS '06, (New York, NY, USA), p.~353–360,
  Association for Computing Machinery, 2006.

\bibitem{process_placement_jeannot}
E.~Jeannot, G.~Mercier, and F.~Tessier, ``Process placement in multicore
  clusters:algorithmic issues and practical techniques,'' {\em IEEE
  Transactions on Parallel and Distributed Systems}, vol.~25, no.~4,
  pp.~993--1002, 2014.

\end{thebibliography}

\newpage
\appendices

\section{Example Transformation of MPI Traces into Execution Graphs}

\begin{figure}[!htp]
\centering
\includegraphics[width=1\linewidth]{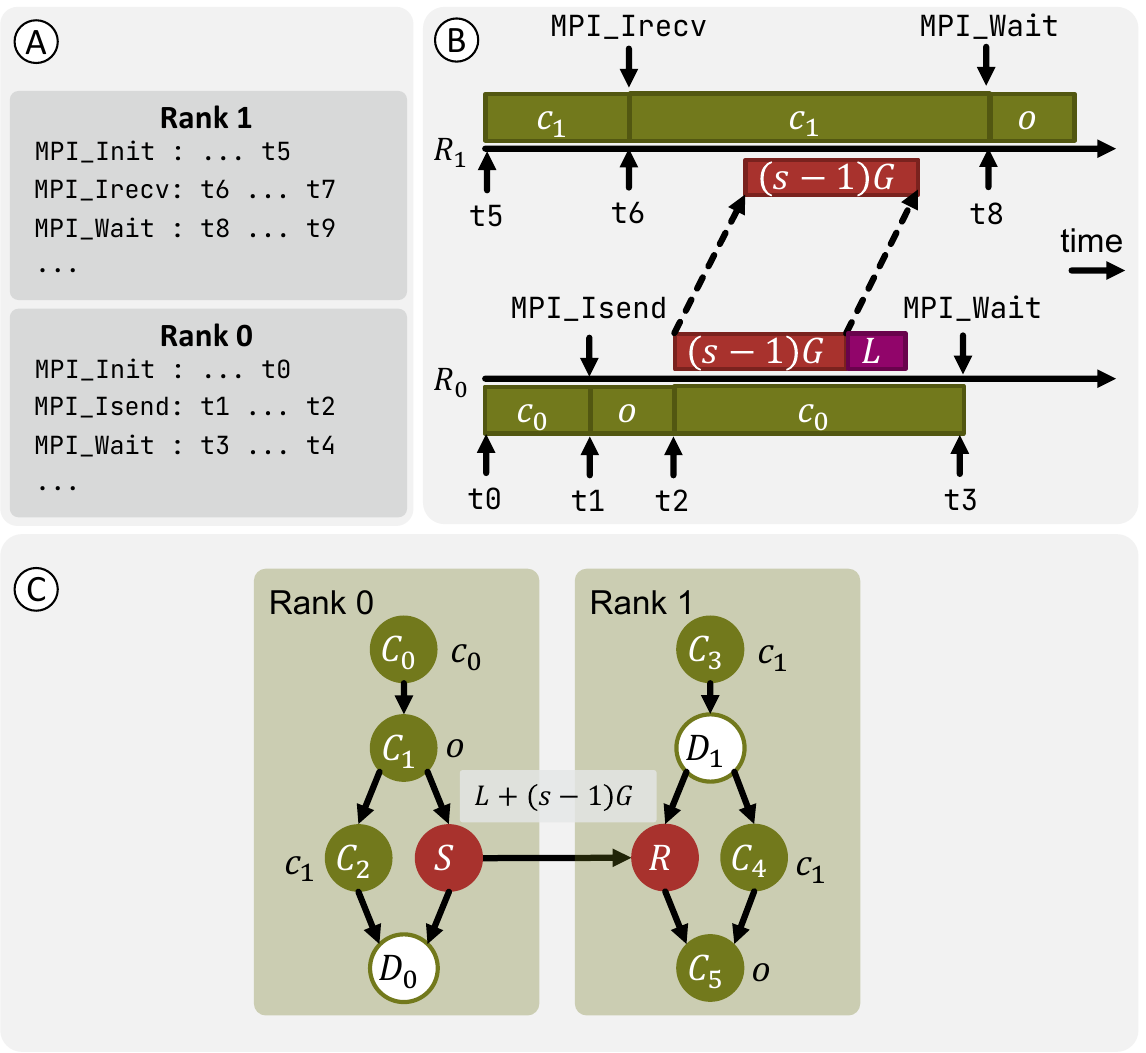}
\cprotect\caption{An example illustrating the transformation of nonblocking p2p operations into execution graphs, under the assumption that the eager protocol is used.}
\label{fig:dependency-schedule-nonblocking-example}
\end{figure}

\section{Support for the Rendezvous Protocol}
\label{appendix:sec:support-for-rendezvous}

\begin{figure}[!htp]
\centering
\includegraphics[width=1\linewidth]{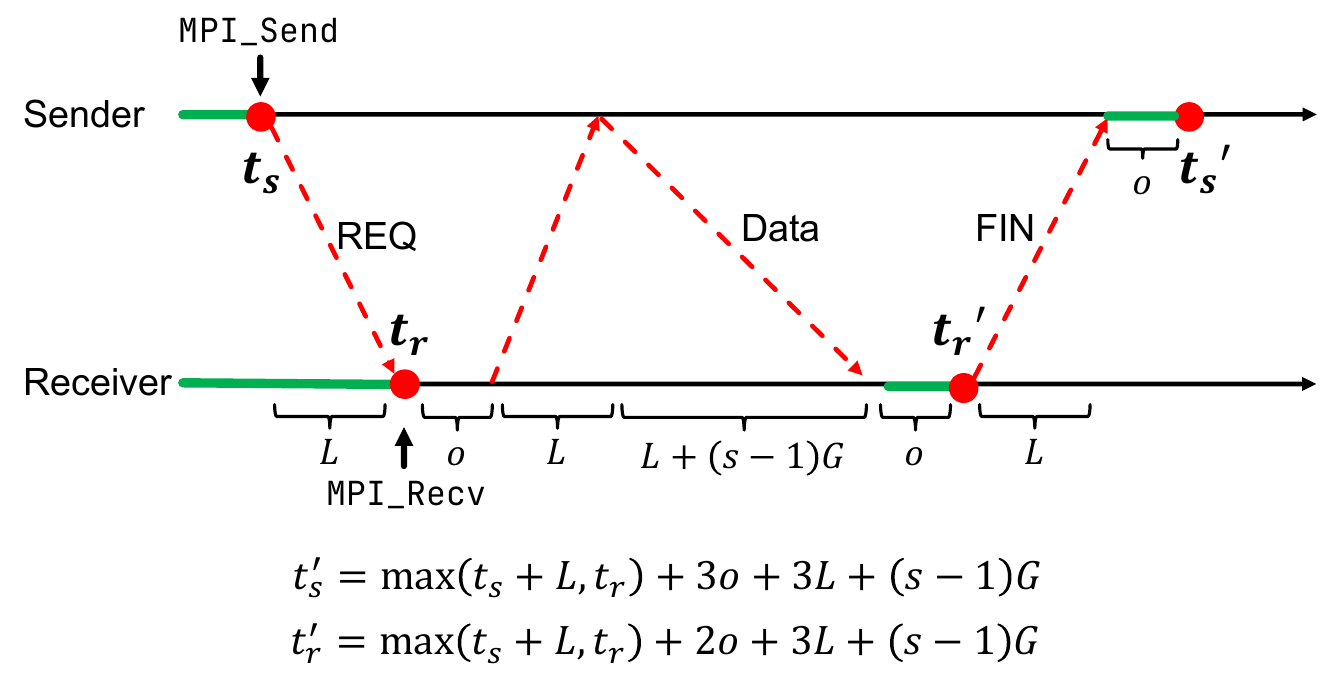}
\cprotect\caption{Space-time diagram illustrating the execution of a blocking p2p MPI operation under the rendezvous protocol with RDMA read-based data transfer.}
\label{fig:loggps-support}
\end{figure}

The LLAMP toolchain fully supports the rendezvous protocol that the original LogGPS paper describes~\cite{loggps_ino}. Fig.~\ref{fig:loggps-support} showcases the execution of \code{MPI_Send} and \code{MPI_Recv} operations within this protocol. Here, $t_s$ and $t_s'$ represent the initiation and completion times of the \code{MPI_Send} operation, respectively, while $t_r$ and $t_r'$ mark the beginning and end of the \code{MPI_Recv} operation. The core concept here is that $t_s'$ and $t_r'$ can be expressed through equations featuring only \textit{max} functions and linear terms, which allows them to be easily converted into linear constraints.

\begin{figure}[!htp]
\centering
\includegraphics[width=.8\linewidth]{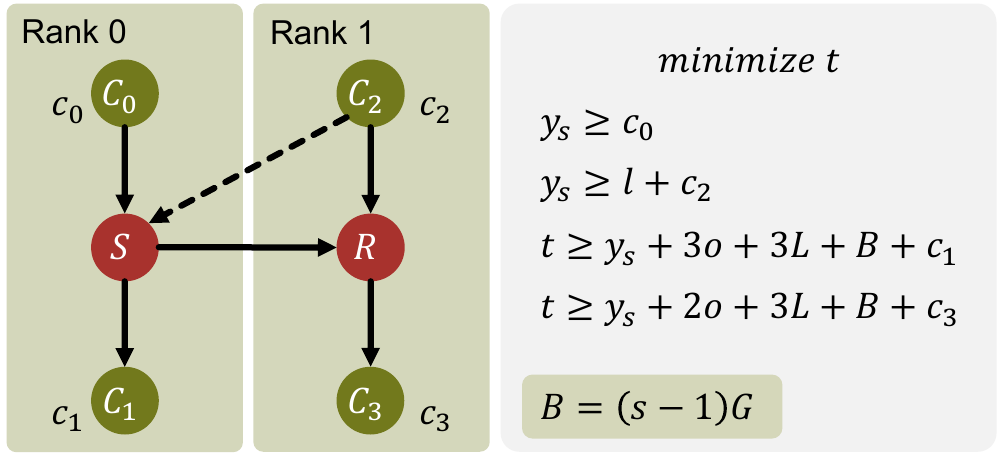}
\cprotect\caption{Linear program that corresponds to the execution graph in Fig.~\ref{fig:dependency-schedule-example}, employing the rendezvous protocol. Assuming the message size is $s$, we simplify the bandwidth cost $(s-1) G$ to $B$ for brevity.}
\label{fig:rendezvous-example}
\end{figure}

In Fig.~\ref{fig:rendezvous-example}, we show the LP generated for the running example when the rendezvous protocol is applied. Note that during the construction of the execution graph, we need to add a virtual edge between $S$ and $C_2$ to reflect that the end time of $S$ is dependent on the start time of $R$. Notably, the final number of constraints matches Equation\ref{eq:lp-model-example}, demonstrating that integrating the rendezvous protocol into the LP does not introduce extra constraints.

\newpage
\section{Linear Program Generation}

\begin{algorithm}[!htp]
\caption{Pseudocode for converting an MPI execution graph into a linear programming problem based on the LogGP model.}
\label{alg:lp-generation}
\DontPrintSemicolon
\SetKwInOut{Input}{Input}
\SetKwInOut{Output}{Output}
\Input{Execution graph $\mathcal{G}$}
\Output{Linear programming model $\mathcal{F}$}
Initialize linear programming model $\mathcal{F}$ \;
Initialize dictionary $T_v$ \;
$\mathcal{l}, \mathcal{g}, \mathcal{o} \gets$ add\_vars($\mathcal{F}$) \; \label{alg:lp-generation:new-vars}
$vs \gets$ topological\_sort($\mathcal{G}$) \; 
\For{each $v$ in $vs$} {
    $preds \gets $ predecessors($v$)\;
    \uIf{$v$ is a starting vertex} {
        $T_v[v] \gets 0$ \;
    }
    \uElseIf{$v$ has only one predecessor} {
        $T_v[v] \gets T_v[preds[0]] +$ cost($v$) \;
    }
    \Else {
        $y \gets $ add\_var($\mathcal{F}$) \;
        \ForEach{$p$ in $preds$} {
            \eIf{$v$ is \textit{recv} and $p$ is \textit{send}} {
                $c \gets (y \geq T_v[p] + \mathcal{l} + (s - 1) \mathcal{g})$ \; \label{alg:lp-generation:comm-cost}
            }{
                $c \gets (y \geq T_v[p])$ \;
            }
            add\_constraint($\mathcal{F}$, $c$) \;
        }
        $T_v[v] \gets y +$ cost($v$) \;
    }
}
set\_objective($\mathcal{F}$, min, $T_v[vs[-1]]$) \;
\Return $\mathcal{F}$
\end{algorithm}

\newpage
\section{Computing critical latencies}
\label{appendix:sec:computing-critical-latencies}

\begin{algorithm}[!htp]
\cprotect\caption{Pseudocode for computing critical latencies within a given interval $[L_{min}, L_{max}]$. The input value $step$ defines the resolution of the algorithm. The \code{SALBLow()} function borrows from the variable attribute with the same name from Gurobi~\cite{gurobi}, and returns the smallest lower bound value at which the current optimal basis would remain optimal.}
\label{alg:critical-latencies}
\DontPrintSemicolon
\SetKwInOut{Input}{Input}
\SetKwInOut{Output}{Output}
\Input{LP model $\mathcal{F}$, variable $\mathcal{l}$, small value $\epsilon$\\
configuration $\theta$, interval $[L_{min}, L_{max}]$, $step$}
\Output{A list of critical latencies $L_{c}$}
Initialize empty list $L_{c}$, tuple $L_{fl}$, variable $\lambda_L$ \;
$L \gets L_{max}$ \;
assign\_var\_lb($\mathcal{F}$, $\theta$) \;
\Repeat{$L_{fl} < L_{min}$}{
    Assign constraint $\mathcal{l} \geq L$ \;
    \tcc{x is the value of the objective value}
    $x$ = optimize($\mathcal{F}$) \;
    $L_{fl}' \gets $SALBLow($\mathcal{l}$) \;
    $\lambda_L' \gets$ reduced\_cost($\mathcal{l})$ \;
    \eIf{$\lambda_L \neq \lambda_L'$}{
        \tcc{If $\lambda_L$ has changed, a new $L_c$ is discovered}
        append($L_{c}$, $L_{fl}'$) \;
        $L_{fl} \gets L_{fl}'$, $\lambda_L \gets \lambda_L'$ \;
    }{
        \tcc{If $\lambda_L$ does not change, update the current lower bound}
        $L_{fl} \gets L_{fl}'$ \;
    }
    $L \gets \min{(L - step, L_{fl}' - \epsilon)}$ \; \label{alg:critical-latencies:min-step}
}
\Return $L_{cs}$
\end{algorithm}

\begin{figure}[!htp]
    \centering
    \includegraphics[width=1\linewidth]{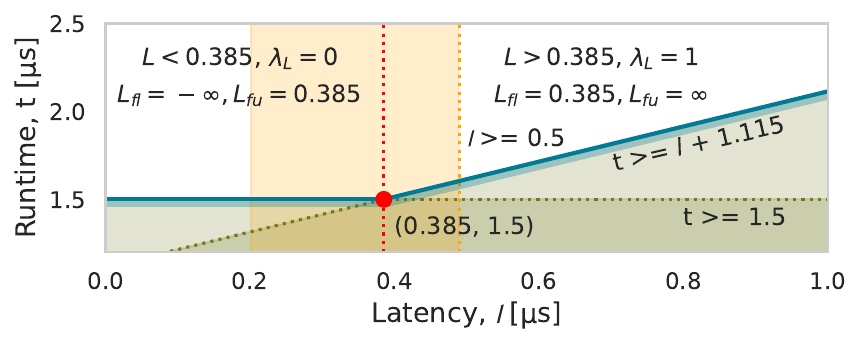}
    \cprotect\caption{Visualization of Algorithm~\ref{alg:critical-latencies} applied to the running example specified by Equation~\ref{eq:lp-model-example}, focusing on the interval between $L_{min} = 0.2\:\mathrm{\mu s}$ and $L_{max} = 0.5\:\mathrm{\mu s}$, depicted by the orange shaded area. The initial lower bound for $l$ is set at $0.5\:\mathrm{\mu s}$. The red line marks the threshold where a basis change occurs, signifying a shift in the critical path within the specified latency range.}
    \label{fig:lp-sen-alg-example}
\end{figure}

As an example, we can execute Algorithm~\ref{alg:critical-latencies} on Equation~\ref{eq:lp-model-example} within an interval $[0.2, 0.5]$. We initially set the lower bound of $l$ to $0.5\:\mu s$, as demonstrated in Fig.~\ref{fig:lp-sen-alg-example}. Upon completing the first iteration, we find $L_{\mathit{fl}} = 0.385\:\mu s$ and $\lambda_L = 1$. We then adjust the lower bound $L$ to $L_{\mathit{fl}} = 0.385 - \epsilon$, ensuring it is slightly below $0.385\:\mu s$ so that the new lower bound falls precisely to the left of $0.385\:\mu s$. This adjustment guarantees that subsequent iterations analyze the region left of the threshold. After two iterations, we successfully identify the critical network latency as $L = 0.385\:\mu s$, and also the network latency sensitivity $\lambda_L$ in each region, showcasing the algorithm's efficiency.

In scenarios involving large LPs, it is common for the basis to change frequently. This means that each iteration of the algorithm reveals only a minimal difference between $L$ and $L_{\mathit{fl}}$, potentially lengthening the time the algorithm requires to run compared to simulators performing parameter sweeps at fixed time intervals. For users seeking a balance between precision and efficiency, rather than identifying every critical latency point within an interval, we introduce the $step$ variable. This variable dictates the minimum shift for the lower bound $L$ after each iteration, ensuring that, in the most time-intensive cases, our algorithm progresses through the interval at a pace comparable to that of simulators operating at a specific time resolution. Given the superior performance of the linear solver as shown in Section~\ref{sec:advantage-of-lp}, this guarantees that LLAMP can always produce results faster than other approaches given the same time resolution.

\section{Experimental Setup and Results for Section \ref{sec:advantage-of-lp}}
\label{appendix:sec:solver-experiment}

All traces were collected in the Ault cluster managed by CSCS. Applications were compiled with GCC 10.2.0, Open MPI 4.1.1. Traces were recorded with liballprof 1.0.

We used the MPI version of NPB~\cite{npb_bailey} 3.4.2, LULESH2.0\footnote{https://github.com/LLNL/LULESH/commit/3e01c40}\cite{lulesh_2_karlin}, and LAMMPS~\cite{lammps_thompson}(August 2023)\footnote{https://github.com/lammps/lammps/commit/7b5dfa2a3b}. LULESH was executed with the parameters \code{-i 1000 -s 32}. LAMMPS was configured for the EAM metallic solid benchmark, using the input script in.eam, and the potential file Cu\_u3.eam.

The runtime comparison experiment was conducted on a single machine equipped with an AMD EPYC 7501 32-Core 2.0 GHz processor and 512 GiB of memory. We utilized Gurobi version 11.0.1 and LogGOPSim version 1.1, and the latter compiled using GCC 10.2.0. Following a procedure similar to that outlined in Algorithm~\ref{alg:critical-latencies}, the experiment required both the linear solver Gurobi and LogGOPSim to iterate over a range of network latency values within a specified interval. For this study, the interval was set between $L_{min} = 3\:\mu s$ and $L_{max} = 13\:\mu s$, with a step size of $1\:\mu s$ for each iteration. The runtimes over these 10 runs were then averaged to obtain the final results. A notable observation was the significant standard deviation in the runtime for LLAMP on the NPB LU benchmark. This variability can be attributed to Gurobi's initial presolve operation, which significantly enhanced the efficiency of subsequent iterations by leveraging the presolve.

\begin{table}[!htp]
\begin{tabular}{@{}ccccc@{}}
\toprule
\textbf{Application} & {\textbf{\makecell{\# Ranks}}} & {\textbf{\makecell{\# Events}}} & \textbf{\makecell{LLAMP\\{[}s{]}}} & \textbf{\makecell{LogGOPSim\\{[}s{]}}} \\ \midrule
\textbf{NPB BT.C}    & 256                      & 48.3 M                    & $36.4 \pm 0.3$              & $248.6 \pm 0.3$                  \\
\textbf{NPB CG.C}    & 256                      & 46.6 M                    & $31.6 \pm 0.1$              & $253.8 \pm 43$                  \\
\textbf{NPB EP.C}    & 256                      & 34.4 K                    & $0.05 \pm 0.004$           & $0.2 \pm0.02$                  \\
\textbf{NPB FT.C}    & 256                      & 5.79 M                    & $6.7 \pm 0.02$              & $886.1 \pm 4.5$                 \\
\textbf{NPB LU.C}    & 256                      & 156 M                     & $296.9 \pm 435$             & $1081.1 \pm 139.6$              \\
\textbf{NPB MG.C}    & 256                      & 5.75 M                    & $4.4 \pm 0.01$              & $32.1 \pm 5.7$                  \\
\textbf{NPB SP.C}    & 256                      & 77.9 M                    & $55.1 \pm 0.3$              & $381.8 \pm 36.6$                \\
\textbf{LULESH}      & 216                      & 61.2 M                    & $56.3 \pm 0.3$              & $338.6 \pm 60.2$                \\
\textbf{LAMMPS}      & 512                      & 11.4 M                    & $9.4 \pm 0.02$              & $65.0 \pm 12.2$                \\ \bottomrule
\end{tabular}
\caption{Runtime comparison of LLAMP using Gurobi compared with LogGOPSim. The third column lists the number of events in the execution graphs generated by Schedgen.}
\label{tab:solver-results}
\end{table}

Converting an execution graph into an LP introduces an overhead, typically under 15 seconds for every 1 million vertices. Despite this initial time investment, solving the LP subsequently is significantly faster than running simulations with alternative tools. Thus, in most scenarios, the benefits of this speedup and the analysis you can perform with an LP justify the initial overhead.

\section{Latency Injector Setup}

\begin{figure}[!htp]
    \centering
    \includegraphics[width=1\linewidth]{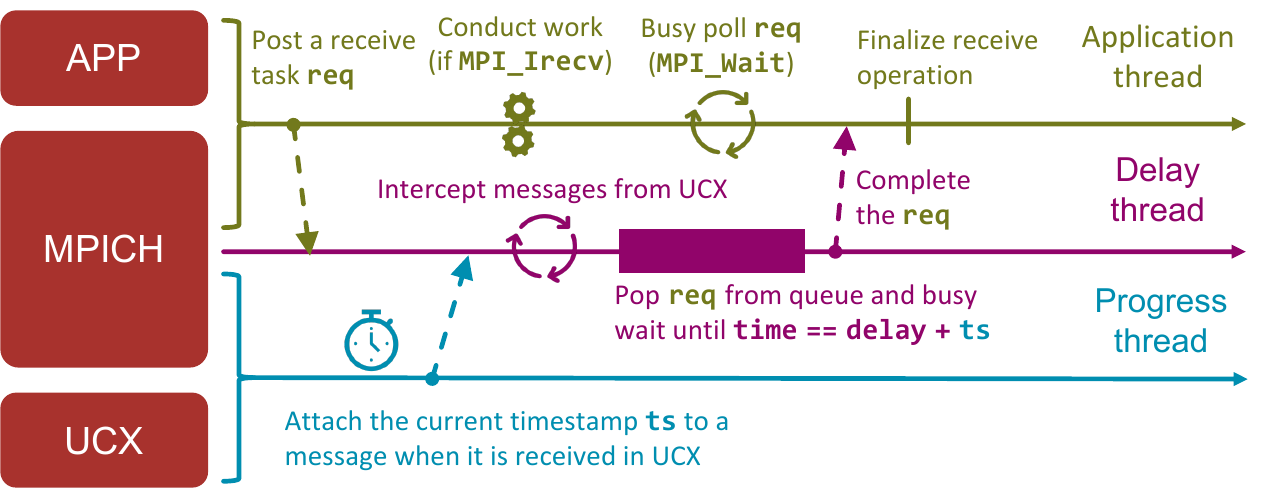}
    \caption{A diagram illustrating the mechanism of our network latency injector built on top of MPICH and UCX.}
    \label{fig:latency-injector-setup}
\end{figure}

\newpage
\section{Validation Experiment}
\label{appendix:sec:validation}

\subsection{LULESH~\cite{lulesh_2_karlin}}

We executed LULESH\footnote{https://github.com/LLNL/LULESH/commit/3e01c40} across all three node configurations using the following arguments
\begin{lstlisting}[language=bash]
    ./lulesh2.0 -i 1000 -s 16
\end{lstlisting}

\subsection{HPCG~\cite{toward_a_new_metric_heroux}}

We executed HPCG\footnote{https://github.com/HPC-benchmark/hpcg/commit/114602d} across all three node configurations using the following arguments:
\begin{lstlisting}[language=bash]
    ./xhpcg 48 48 48
\end{lstlisting}

\subsection{MILC~\cite{milc_bernard}}

MILC\footnote{https://github.com/milc-qcd/milc\_qcd/commit/f03f531} was compiled QIO support from SciDAC. In this work, we focus on the su3\_rmd application of MILC.
The lattice file \code{16x16x16x16.chlat} we used for all the experiments can be obtained from the following site:
\url{https://portal.nersc.gov/project/m888/apex/MILC_lattices/}.

\subsection{ICON\cite{icon_zangl}}

In this work, we used an internal version of ICON, namely 2.6.7. It was compiled with netcdf 4.7.4, netcdf-fortran 4.6.1, and OpenBLAS 0.3.26.dev. The run script, which we include in the artifact is based on the experiment named \code{aquaplanet_04}. The grid file we used is coded \code{R02B04_0013}, which has a resolution of 160 km.

Note that the same configuration was also applied to the case study. The distinction lies in the forecast duration: the validation experiments targeted a 6-hour weather forecast, while the case study extended the forecast to 24 hours.

\subsection{LAMMPS\cite{lammps_thompson}}

For LAMMPS\footnote{https://github.com/lammps/lammps/commit/27d065a}, we ran the EAM metallic solid benchmark for all node configurations, using the input script in.eam, and the potential file Cu\_u3.eam. To minimize runtime variability due to random initial conditions of atoms, we changed the velocity setting to \code{velocity all set 0.1 0.2 0.3}, and the neighbor list setting to \code{neigh_modify once yes}. We tested LAMMPS for weak scaling, and in each node configuration, we made sure that there were 256000 atoms per MPI rank.

\subsection{OpenMX\cite{openmx_boker}}

We used OpenMX 3.7\footnote{https://www.openmx-square.org/openmx3.7.tar.gz} for our experiment, and ran the bulk diamond example (DIA64\_DC) across all node configurations.

\subsection{CloverLeaf\cite{cloverleaf_mallinson}}

We used Cloverleaf 1.3\footnote{https://github.com/UK-MAC/CloverLeaf\_ref/commit/0fdb917}. The input we used is exactly the same as the default input file except that we changed \code{tiles_per_chunk} to 50 and \code{end_step} to 150.

\begin{table}[!htp]
\centering
\begin{tabular}{@{}cccccc@{}}
\toprule
\textbf{Application}                          & {\textbf{\makecell{\# Proc /\\\# Nodes}}} & \textbf{o {[}$\mu s${]}} & \textbf{\# Events} & {\textbf{\makecell{RMSE\\{[}s{]}}}} & {\textbf{\makecell{RRMSE\\{[}\%{]}}}} \\ \midrule
\multirow{3}{*}{\textbf{LULESH}}              & 128/8                       & 5.0                      & 895 K              & 0.031                 & 0.54                    \\
                                              & 432/27                      & 5.0                      & 4.55 M             & 0.038                 & 0.61                    \\
                                              & 1024/64                     & 4.0                      & 13.5 M             & 0.037                 & 0.58                    \\ \midrule
\multirow{3}{*}{\textbf{HPCG}}                & 128/8                       & 5.6                      & 738 K              & 0.053                 & 1.02                    \\
                                              & 512/32                      & 5.0                      & 4.83 M             & 0.099                 & 1.42                   \\
                                              & 1024/64                     & 5.0                      & 12.2 M             & 0.095                 & 1.10                     \\ \midrule
\multirow{3}{*}{\textbf{\makecell{MILC\\{[}su3\_rmd{]}}}} & 128/8                       & 6.0                      & 2.79 M             & 0.052                 & 0.60                    \\
                                              & 512/32                      & 6.0                      & 12.1 M             & 0.050                 & 1.05                    \\
                                              & 1024/64                     & 6.0                      & 23.6 M             & 0.068                 & 1.69                    \\ \midrule
\multirow{3}{*}{\textbf{ICON}}                & 128/8                       & 20.0                     & 362 K              & 0.074                 & 0.34                    \\
                                              & 512/32                      & 16.0                     & 1.89 M             & 0.051                 & 0.52                    \\
                                              & 1024/64                     & 8.6                      & 3.82 M             & 0.053                 & 0.75                    \\ \midrule
\multirow{3}{*}{\textbf{LAMMPS}}              & 128/8                       & 32.4                     & 155 K              & 0.083                 & 1.40                    \\
                                              & 512/32                      & 32.7                     & 596 K              & 0.102                 & 1.46                    \\
                                              & 1024/64                     & 31.7                     & 1.21 M             & 0.128                 & 1.65                    \\ \midrule
\multirow{2}{*}{\textbf{OpenMX}}              & 128/8                       & 15.6                     & 271 K              & 0.204                 & 0.88                    \\
                                              & 512/32                      & 10.9                     & 1.78 M             & 0.097                 & 0.55                    \\ \midrule
\textbf{CloverLeaf}                           & 128/8                       & 6.1                      & 162 K              & 0.037                 & 0.70                   \\ \bottomrule
\end{tabular}
\caption{Validation results across all chosen applications. The fourth column lists the number of events in the execution graphs generated by Schedgen.}
\label{tab:full-validation-results}
\end{table}

\newpage
\section{Support for Network Topology Analysis}

\begin{figure}[!htp]
    \centering
    \includegraphics[width=.8\linewidth]{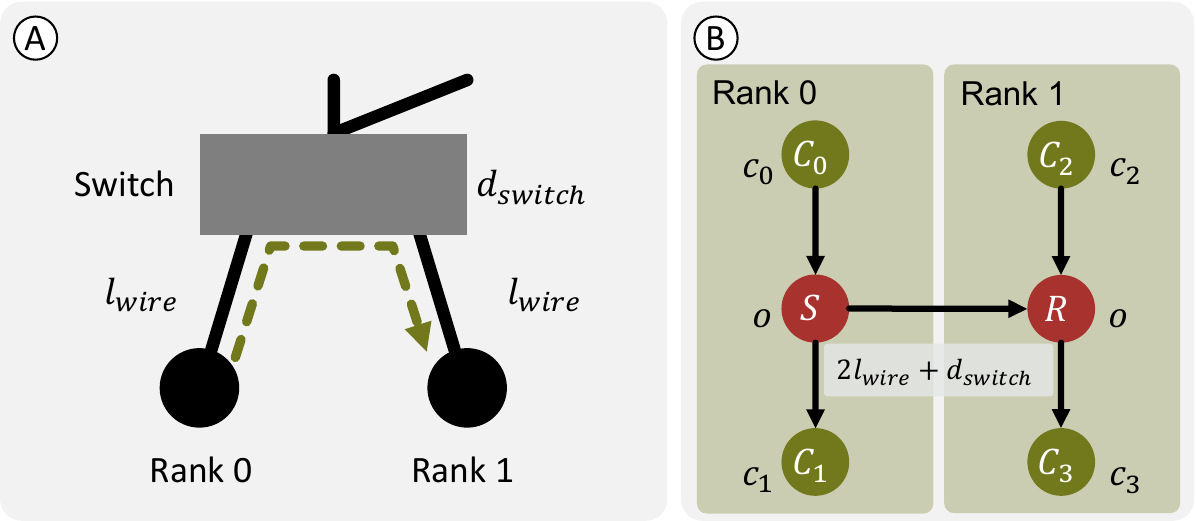}
    \caption{An example demonstrating that by replacing the cost of communication edges of the execution graph with appropriate decision variables, we can investigate the impact of the wire latency as well as the network topology on the overall performance of an application.}
    \label{fig:topology-example}
\end{figure}

\begin{figure}[!htp]
    \centering
    \includegraphics[width=.8\linewidth]{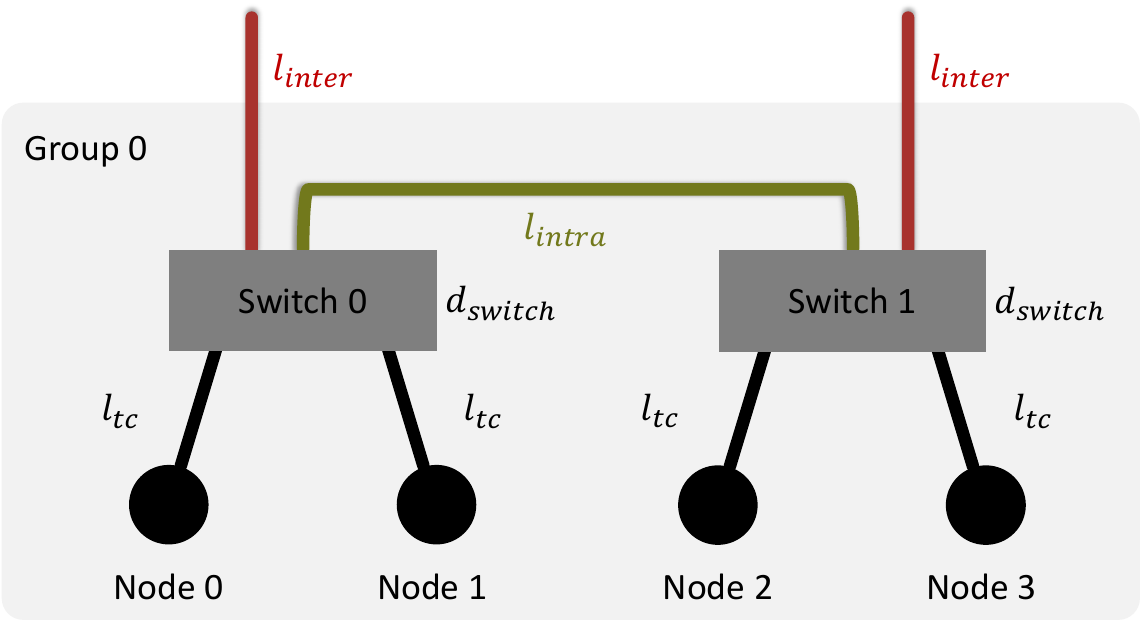}
    \caption{Illustration of a simplified Dragonfly network topology, highlighting that different types of wires can be represented using unique decision variables for variable latencies.}
    \label{fig:dragonfly-topology-example}
\end{figure}

LLAMP is fully capable of modeling network topologies with different link latencies, as illustrated in Fig.\ref{fig:dragonfly-topology-example}. This figure presents a section of a Dragonfly network topology where different latencies are assigned to specific types of connections. The latency from the switch to a node is denoted by $l_{\mathrm{tc}}$, with $tc$ referring to a \emph{terminal channel}, a term from the original Dragonfly paper\cite{dragonfly_kim}. Latencies for intra-group switch connections are assigned to $l_{\mathrm{intra}}$, while inter-group link latencies are marked as $l_{\mathrm{inter}}$. To assess the tolerance for a particular type of link, one simply sets the lower bounds of other variables as constants and conducts optimization as detailed in Section~\ref{sec:network-latency-tolerance}. For instance, to find the maximum tolerable latency of an inter-group channel ($l_{\mathrm{inter}}$) in a Dragonfly topology, one sets $l_{\mathrm{tc}} \geq L_{\mathrm{tc}}$ and $l_{\mathrm{intra}} \geq L_{\mathrm{intra}}$, where $L_{\mathrm{tc}}$ and $L_{\mathrm{intra}}$ are fixed constants, and then solves for $\max l_{\mathrm{inter}}$. 
This process enables the derivation of all performance metrics via identical optimization steps one variable at a time.

\section{Support for Heterogeneous LogGP}
\label{appendix:sec:heterogenenous-loggp}
One limitation of the LogGPS model is that it assumes applications are run on homogeneous single-processor nodes linked by a single network with uniform latency and bandwidth. In the context of process mapping, heterogeneity needs to be considered since, for instance, the intra-node communication latency will be significantly lower than inter-node communication latency. To incorporate heterogeneity, we re-define the $L$ and $G$ parameters as matrices of size $P \times P$, where elements $L_{i,j}$ and $G_{i,j}$ represent the latency and bandwidth between ranks $i$ and $j$, respectively. These matrices are symmetric as we assume symmetric communication costs between rank pairs. This modified model aligns with a simplified HLogGP model developed by Bosque et al.~\cite{hloggp_bosque}. Since we focus on the impact of the network, we only introduce different $L$ and $G$ and assume that all the other parameters, namely $o$, $g$, and computational power, will be the same across all ranks. Hence, we define \emph{pair-wise network latency sensitivity} as
\begin{align}
\lambda_L^{i,j}(\mathcal{G}, \theta, \Phi, \pi) = \partial T(\mathcal{G}, \theta, \Phi, \pi) / \partial L_{i,j}
\end{align}
where $\Phi$ is an architecture topology graph containing the latency and bandwidth between each processor pair, and $\pi$ is a specific process mapping that assigns each rank to a processor. Similarly, the pair-wise bandwidth sensitivity can be expressed as $\lambda_G^{i,j}(\mathcal{G}, \theta, \Phi, \pi) = \partial T(\mathcal{G}, \theta, \Phi, \pi) / \partial G_{i,j}$.

When constructing the LP model, $\mathcal{l}$ and $\mathcal{g}$ are replaced by the corresponding $\mathcal{l}_{i,j}$ and $\mathcal{g}_{i,j}$, namely on lines \ref{alg:lp-generation:new-vars} and \ref{alg:lp-generation:comm-cost} in Algorithm~\ref{alg:lp-generation}. Consequently, after assigning lower bounds to each $\mathcal{l}_{i,j}$ and $\mathcal{g}_{i,j}$ based on a given architectural topology and solving the LP model, we will be able to acquire pair-wise sensitivity measures for both $L$ and $G$ by simply reading the reduced cost of the corresponding decision variable.

\section{Rank Placement}
\label{appendix:sec:rank-placement}

Rank placement is crucial for achieving optimal performance in parallel applications. It involves determining the mapping of MPI ranks onto physical processors on computational nodes. The goal is to balance the workload and communication among ranks, and minimize communication overhead. The established state-of-the-art approach usually begins by profiling the communication pattern (i.e., the number of bytes transferred between each pair of ranks). Subsequently, software tools, such as graph partitioners, are employed to compute the optimal process mapping~\cite{process_mapping_zhang, mpipp_chen, process_placement_jeannot}. However, these methods only focus on the communication volume and disregard the temporal characteristics of applications. In this section, we present a novel iterative heuristic that leverages sensitivity measures and critical path analysis, facilitated by LP, to compute process mapping.

\begin{algorithm}[!htp]
\caption{Pseudocode for process mapping}\label{alg:process-mapping}
\DontPrintSemicolon
\SetKwInOut{Input}{Input}
\SetKwInOut{Output}{Output}
\Input{Linear programming model $\mathcal{F}$,\\architecture graph $\Phi$}
\Output{Process mapping $\pi$}
$\pi \gets $ generate\_initial\_mapping($\Phi$) \;
$f^* \gets \infty$ \;
\While{$!stop$} {
    \tcc{Assign lower bounds to $\mathcal{l}_{i,j}$ and $\mathcal{g}_{i,j}$}
    assign\_var\_lb($\mathcal{F}$, $\pi$, $\Phi$) \;
    optimize($\mathcal{F}$) \;
    \tcc{Compute new objective value}
    $f \gets $ objective\_val($\mathcal{F}$) \;
    \eIf{$f < f^*$} {
        $f^* \gets f$ \;
    }{
        Revert back to the previous $\pi$ \;
        \tcc{Terminate if $f$ does not improve} 
        $stop \gets true$ \;
    }
    $\mathcal{D}_{L}$, $\mathcal{D}_{G} \gets $
    get\_sensitivity\_matrices($\mathcal{F}$) \;
    \ForEach{pair of ranks $i$, $j$ in $\pi$} {
        $gain \gets$ swap\_gain($i$, $j$, $\mathcal{D}_L$, $\mathcal{D}_G$, $\pi$, $\Phi$)\;
        Keep track of $i^*, j^*$ that yield the highest $gain$ \;
    }
    \eIf{Cannot find $i$ and $j$ that yield positive $gain$} {
        $stop \gets true$ \;
    }{
        Swap rank $i^*$ and $j^*$ in $pi$ \;
    }
}
\Return $\pi$
\end{algorithm}

The primary idea of the algorithm is that we use an iterative algorithm to gradually refine an initial mapping that can be either randomly generated or given by the user. In each iteration, we first optimize the linear program and obtain the latency and bandwidth sensitivity matrices, which contain information about the number of messages and the total data transferred between each pair of ranks along the critical path. Then, based on these matrices, we compute the potential performance gain by swapping each pair of ranks in the current mapping. At the end of the iteration, we swap the ranks that will likely yield the most performance gain in the next iteration. Since the objective value represents the estimated runtime of the program, we can accurately determine whether the swaps improve the performance of the application. The two termination conditions are: first, when we cannot find another pair of ranks to swap that can potentially yield positive performance gain; second, when the estimated runtime (i.e., the objective value of the function) is worse after a swap.

\subsection{Preliminary Results}

\begin{figure}[!htp]
    \centering
    \includegraphics[width=.9\linewidth]{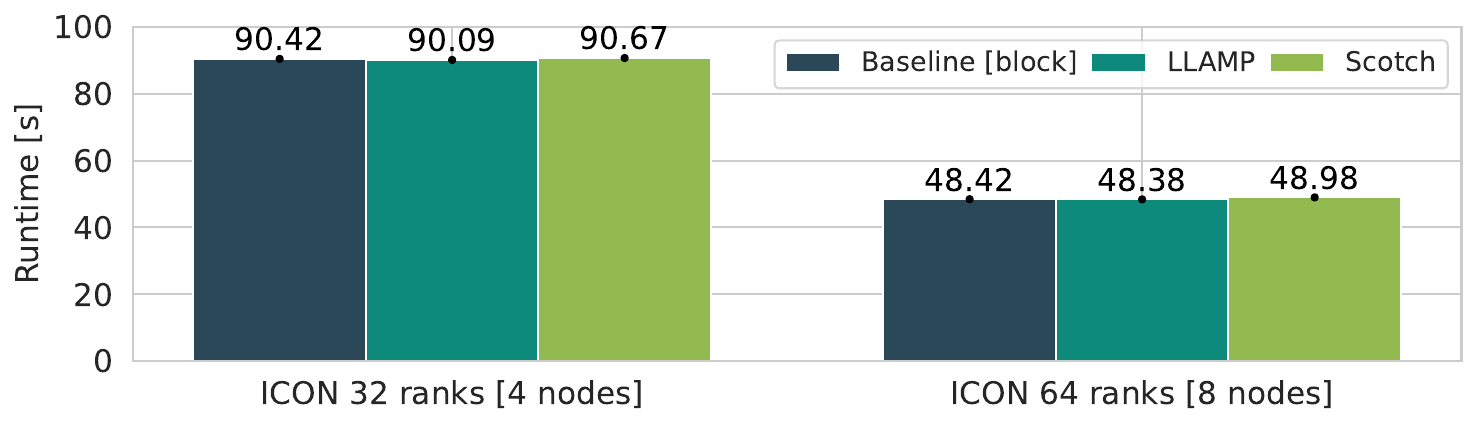}
    \caption{Preliminary results of our rank placement algorithm compared with the default option (block) and Sctoch.}
    \label{fig:icon-process-map}
\end{figure}

This section outlines initial findings related to LLAMP's rank placement algorithm. We compare our approach with the default scheme (\emph{block}) and the results obtained using Scotch, a state-of-the-art software that computes static process mappings through traffic volume profiling between rank pairs, following the methodology of Jeannot et al.~\cite{process_placement_jeannot}.

The evaluation was conducted on the Piz Daint supercomputer that uses Cray MPICH. We present the outcome of preliminary trials carried out on the ICON model, using 32 ranks across 4 nodes and 64 ranks on 8 nodes, as shown in Fig.~\ref{fig:icon-process-map}. Each configuration was executed 10 times and we report the average runtime. The results show that our rank placement algorithm resulted in a slight reduction in ICON's runtime. However, the improvement is so minor that it might be attributed to normal variations in system and network performance. Scotch's performance was slightly worse, potentially because ICON has already been heavily optimized and Scotch does not take into account the temporal behavior of the application, focusing only on data volume, which may not be the most critical factor for this application. Nonetheless, since our algorithms only yielded less than 1\% improvement, the results are inconclusive.

This investigation is an initial step in exploring one of the many potential applications of the LLAMP toolchain and our new approach employing linear programming. It is by no means an exhaustive analysis or comparison of all rank placement algorithms available. Its purpose is to showcase the flexibility of our toolchain.
\end{document}